# Title: Thermodynamic phase transitions reveal the resilience structure of urban traffic congestion

Luis E. Olmos

Universidad EAFIT, Medellín, Colombia

**Abstract:** Understanding how cities transition from free-flowing to congested traffic remains a central open problem in urban science. Here we show that city-scale congestion undergoes a reproducible nonlinear transition analogous to an order-disorder phase transition in statistical mechanics, in which aggregate mobility acts as a control parameter and jam extent as a collective order parameter. Crucially, this analogy is not merely formal: we derive and empirically identify an effective thermodynamic temperature with concrete physical meaning, quantifying infrastructural heterogeneity and how broadly a city explores congestion configurations as demand increases. Low-temperature cities are congestion-fragile: small mobility increases trigger sharp, system-wide jam transitions. This framework further reveals that the macroscopic fundamental diagram is an incomplete description of the traffic state: it emerges as a projection of a richer free-energy landscape governed by entropy-capacity trade-offs. Validated across 46 cities in Latin America and the Caribbean and independently confirmed with loop-detector data from 8 cities on three continents, these results establish a physics-based foundation for comparing urban traffic resilience and anticipating congestion regime shifts under changing mobility demand.

## Introduction

Urban traffic congestion exhibits a fundamental paradox: despite large daily fluctuations in travel demand, cities repeatedly organize into similar macroscopic traffic states, yet no unified framework explains why cities with comparable infrastructure respond so differently to the same mobility load. Early work by Prigogine and Herman (1) recognized the collective nature of traffic and called explicitly for a statistical-mechanical description linking microscopic driver behavior to city-scale flow patterns — a program that, nearly five decades later, remains incomplete.

Subsequent research has revealed important macroscopic regularities: reproducible relations between flow, density, and speed captured by the macroscopic fundamental diagram (MFD) (2–6), well-defined network capacity limits (7–10), and percolation-like congestion transitions (11–13). These findings establish congestion as an emergent network-level phenomenon. Yet the MFD, despite its empirical robustness, remains a descriptive tool: it does not explain what determines the shape of the transition between free-flow and congested regimes, why saturation occurs at city-specific mobility levels, or how infrastructural heterogeneity and demand combine to produce stable global responses (2,5,7,14,15). A principled framework distinguishing control parameters from collective responses at the city scale is still lacking (16).

Here we close this gap by resolving a long-standing open problem in traffic science. Thermodynamic analogies for vehicular traffic have been proposed for decades (1,14,15), but have remained purely formal: no study has identified what the effective temperature actually is in terms of measurable urban properties, nor derived it from first principles. Using mobility and congestion data from 46 metropolitan areas across Latin America and the Caribbean — where the COVID-19 pandemic provided an unplanned natural experiment spanning traffic regimes rarely accessible under normal conditions — we identify mobility as a control parameter and jam extent as a collective order parameter of the urban traffic state. We derive and empirically identify the effective thermodynamic temperature as a concrete structural property of the city, determined by infrastructural heterogeneity and network availability, which governs a city's resilience to congestion: low-temperature cities undergo sharp, system-wide jam transitions under small mobility increases, while high-temperature cities absorb demand gradually. This is not an analogy — it moves the thermodynamic framework toward a quantitative, empirically grounded identification of urban traffic states.

This framework further reveals that the macroscopic fundamental diagram is an incomplete description of the traffic state. Rather than a primitive empirical law, the MFD emerges as a collective envelope shaped by entropy-capacity trade-offs, a projection of a richer free-energy landscape in which the appropriate macroscopic object is the effective Helmholtz free energy. Validated independently with loop-detector measurements from 8 cities on three continents, these results establish a unified, physics-based foundation for understanding urban traffic resilience and anticipating regime shifts under changing mobility demand.

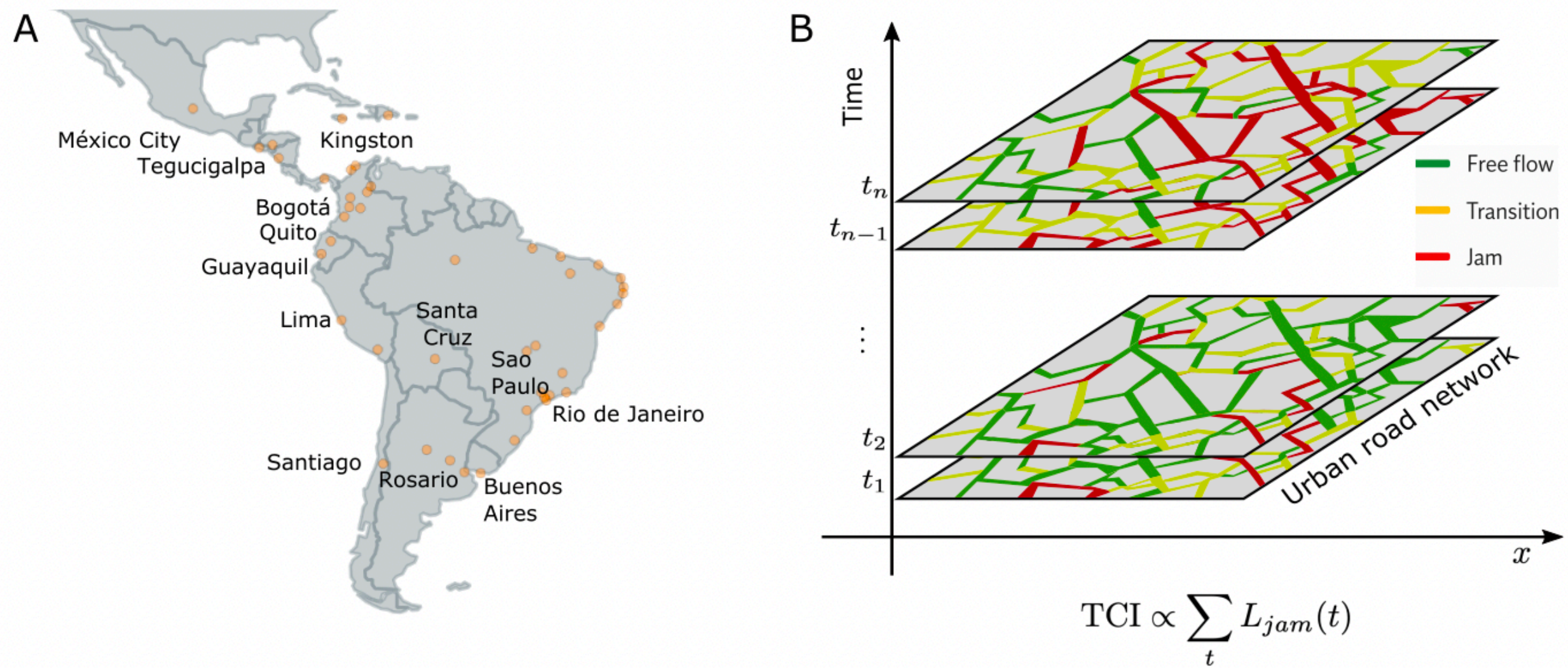

**Figure 1**. **Dataset coverage and conceptual framework for city-scale congestion measurement.** (A) Geographic distribution of the 46 metropolitan areas analyzed across Latin America and the Caribbean. The dataset spans a wide range of city sizes, population densities, and infrastructure levels, providing a regionally consistent sample for cross-city comparison. (B) Conceptual illustration of the Traffic Congestion Index (TCI) (16,17). At each 5-minute snapshot, road segments are classified as free-flowing (green), transitioning (yellow), or jammed (red) based on observed speeds relative to free-flow conditions. The lengths of all jammed segments are summed across the network at each instant and accumulated over the full day, yielding a structural measure of congestion that captures both spatial extent and temporal persistence. The daily TCI is proportional to this time-aggregated jam length, $TCI \propto \sum_t L_{jam}(t)$ and is expressed as a percentage change relative to a pre-pandemic baseline, enabling consistent comparison across cities.

## Results

### Aggregate mobility drives a reproducible phase transition in city-scale congestion

City-scale traffic states can be characterized by two macroscopic observables: jam extent and aggregate human mobility. Congestion is quantified using the Traffic Congestion Index (TCI) (16,17), a dimensionless daily indicator that aggregates the total length of congested road segments across the network and over time, capturing both spatial extent and temporal persistence of jams (Fig. 1B). Mobility demand is measured from workplace visits reported in Google Community Mobility Reports (18), a proxy for aggregate commuting demand. Both variables are expressed relative to pre-pandemic baselines, enabling consistent comparison across the 46 metropolitan areas analyzed across Latin America and the Caribbean (Fig. 1A; see Methods and Supplementary Note 1 for details).

Across cities, congestion responds to mobility through a reproducible nonlinear transition. In Rio de Janeiro, the relationship is well described by a sigmoidal function (Fig. 2A): congestion grows gradually at low mobility, rises sharply over an intermediate range, and saturates at high demand levels — the hallmark of a collective phase transition between predominantly free-flow and predominantly congested network states. This pattern is not idiosyncratic. A sigmoidal fit describes the mobility–congestion relation with $R^2 > 0.8$ in 31 of the 46 cities analyzed (SI Fig. S2), and moderately well ($0.7 < R^2 \leq 0.8$) in eight additional cities (SI Fig. S3), together accounting for 85% of the dataset. The main exceptions are Brazilian cities affected by prolonged social disruptions, where heterogeneous mobility adherence weakened the emergence of a well-defined transition (SI Fig. S3). The prevalence and consistency of this pattern across diverse urban contexts — spanning different sizes, densities, and infrastructure levels — establishes mobility as an effective control parameter of macroscopic traffic states.

The sigmoidal transition is not an artifact of normalization. In cities with low baseline congestion, the saturation plateau is not reached even when mobility returns to pre-pandemic levels, reflecting intrinsic network constraints rather than a ceiling imposed by the baseline procedure. Several cities show post-pandemic congestion increases beyond 100% mobility without reaching saturation, confirming that the transition threshold is city-specific and governed by structural network properties rather than by the normalization scheme.

Crucially, this transition is unique to the TCI and cannot be recovered from other standard congestion metrics. Travel-time indices (19) increase approximately linearly with mobility (Fig. 2B), tracking aggregate demand without capturing the collective reorganization of network states, because averaging over heterogeneous travel experiences smooths out the spatial clustering and buildup of jams that defines the transition. Vehicle miles traveled (20) saturate at high mobility but without a sigmoidal signature (Fig. 2C), suggesting that while network capacity limits total driving activity, it does not produce the sharp collective reorganization captured by the TCI. The TCI is distinctive precisely because it measures the fraction of the network structurally occupied by congestion; a collective configuration of the road system rather than an outcome of individual travel experiences. This structural character identifies it as the natural order parameter for city-scale traffic transitions.

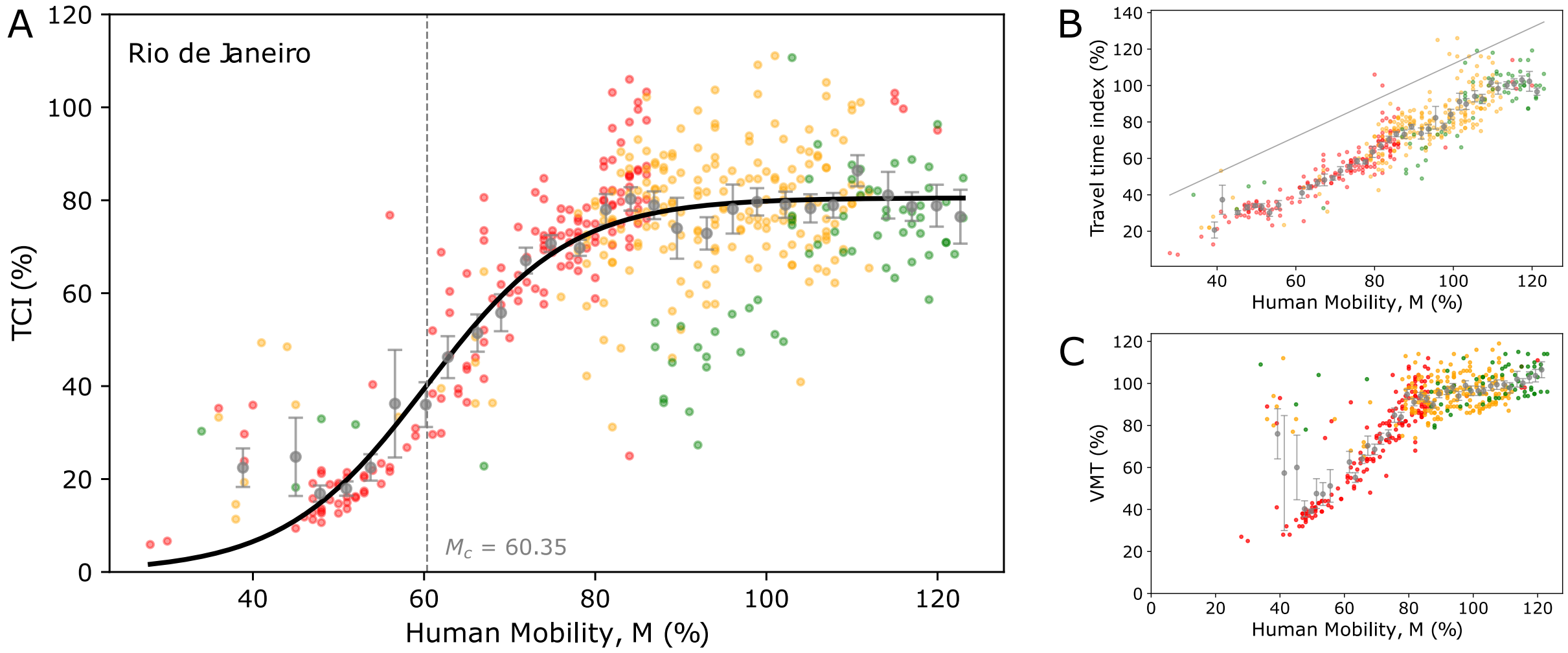

**Fig. 2. Mobility drives a reproducible phase transition in city-scale congestion.** Shown for Rio de Janeiro as an illustrative example; similar patterns are observed across 39 of 46 cities analyzed (SI Appendix, Figs. S2–S5). (A) Daily normalized Traffic Congestion Index (TCI) versus daily human mobility, excluding weekends and major disruption days. Each point represents one day, colored by year (2020, red; 2021, yellow; 2022, green). Gray points and error bars indicate binned means and variability across mobility intervals. The solid curve shows a sigmoidal fit — the hallmark of a collective phase transition — and the vertical dashed line marks the characteristic mobility level $M_c$ at which congestion reaches half its saturation value. The persistence of this sigmoidal shape across three years spanning vastly different mobility regimes establishes mobility as an effective control parameter of the macroscopic traffic state. (B) Daily travel-time index reported by the TomTom Traffic Index (20) versus mobility. The approximately linear response — indicated by the diagonal reference line (slope = 1) — contrasts sharply with the nonlinear transition in panel A, revealing that travel-time indices track aggregate demand rather than the collective reorganization of network states. (C) Daily vehicle miles traveled (VMT) versus mobility(20) . VMT increases approximately linearly at low-to-intermediate mobility and saturates at high demand, but without the sigmoidal signature of panel A — confirming that capacity constraints alone do not produce the sharp collective transition captured by the TCI.

## A thermodynamic framework describes congestion transitions across cities

The robust sigmoidal mobility–congestion relation observed across cities suggests that city-scale traffic can be described as a collective transition between predominantly uncongested and predominantly congested network states. At the macroscopic level, the TCI is proportional, up to normalization, to the fraction of congested road segments, providing a direct measure of the global congestion state of a city. Within this coarse-grained picture, congestion emerges as the collective response of many binary units to increasing mobility demand: daily mobility acts as an effective external field (21,22) favoring congested configurations, while the TCI summarizes the macroscopic response of the network.

This binary description admits a minimal free-energy formulation. As shown in Supplementary Note 4, treating road segments as independent binary units — analogous to spins in a noninteracting Ising-like system in an external field (23) — equilibrium minimization of the resulting free energy yields a hyperbolic-tangent equation of state, which at the city scale describes the mobility–congestion relation. At the city scale, mobility M plays the role of an effective control variable analogous to network density — both drive the system toward congested configurations — while the TCI provides the corresponding macroscopic order parameter. Under this mapping, the same functional form describes the mobility–congestion relation. After centering mobility by the city-specific transition point $M_c$— defined as the mobility level at which congestion reaches half its empirical saturation value — the mobility–congestion relation takes the form

$$2\langle m\rangle - 1 = \tanh\left(\frac{M-Mc}{T}\right), \qquad (1)$$

where $\langle m\rangle$ denotes the TCI normalized by its empirical saturation value, mapping congestion onto the symmetric interval $[-1,1]$ consistent with the binary classification of road segments. The parameter T acts as an effective temperature controlling the width of the transition: lower values correspond to sharper mobility–congestion responses, while larger values produce broader, more gradual transitions. Eq. 1 is not imposed by analogy — it reflects a coarse-grained thermodynamic description of urban traffic that requires no assumptions about microscopic origin-destination patterns or individual route choice.

Figure 3A shows the fitted curves for multiple cities, revealing systematic cross-city differences captured by this single parameter. The collapse of diverse cities onto a common functional form , differing only in $T$ and $M_c$, is itself a nontrivial result: it establishes that the macroscopic congestion response is governed by two city-specific parameters, one structural ($T$) and one demand-related ($M_c$), regardless of the details of urban form or travel behavior.

This formulation admits a full thermodynamic interpretation. Eq. 1 is the derivative of an effective Helmholtz free energy with respect to mobility,

$$2\langle m\rangle - 1 = -\left(\frac{\partial F}{\partial M}\right)_T , \qquad (2)$$

where

$$F(T,M) = \langle E\rangle - TS = -T\ln\left[2\cosh\left(\frac{M-M_c}{T}\right)\right]. \qquad (3)$$

This is the canonical free energy of noninteracting binary units in an external field. Variations in mobility reshape the effective free-energy landscape: its first derivative determines the sensitivity of congestion to mobility, while its curvature controls the sharpness of the transition. Figure 3B shows the reconstructed free-energy profiles for all analyzed cities. Cities with lower effective temperature exhibit narrower profiles around the transition — reflecting more abrupt collective responses — while higher-temperature cities display broader, smoother landscapes. A complete summary of fitted parameters is provided in Supplementary Table S1.

**Effective temperature as a measurable indicator of urban traffic resilience**

The effective temperature $T$ inferred from the canonical fits is not merely a fitting parameter — it is a measurable structural property of the city that determines its resilience to traffic regime shifts. Cities with low $T$ are congestion-fragile: small increases in mobility trigger sharp, system-wide transitions toward saturated congestion, and the network's collective response is dominated by a limited subset of structurally critical corridors that recurrently absorb demand. Cities with high $T$ are congestion-resilient: they respond gradually to increasing mobility, distributing congestion more broadly across the network and absorbing demand without abrupt collective reorganization. The effective temperature thus captures a fundamental asymmetry between cities that is invisible to standard congestion metrics: two cities can have identical average congestion levels yet differ dramatically in how abruptly they approach saturation.

In thermodynamic terms, temperature quantifies how rapidly the number of accessible configurations changes with energy, or equivalently, how broadly the system explores its configuration space as external forcing increases. In urban traffic, the road network defines the available configuration volume, road segments provide the degrees of freedom, and mobility acts as the external field biasing the system toward congested states. Within this picture, $T$ quantifies the configurational flexibility of the network: how broadly congestion patterns are redistributed across the city as demand grows. Low-temperature cities explore a narrow, recurrent set of congestion configurations; high-temperature cities explore a wider, more distributed landscape.

This interpretation is directly supported by empirical relationships with measurable urban properties (Fig. 3C–E). The effective temperature decreases systematically with population density (Fig. 3C), as denser cities have less configurational flexibility per unit demand, and increases with road network length per capita (Fig. 3D), as cities with more infrastructure per person have more degrees of freedom available to absorb and redistribute congestion. Most directly, $T$ is negatively correlated with the annual congestion index reported by TomTom (Fig. 3E): cities with lower effective temperature are systematically more congestion-prone, reaching saturation at lower relative mobility and undergoing sharper regime changes. These three relationships — spanning infrastructure, density, and chronic congestion — confirm that T captures the combined influence of spatial organization and infrastructural availability on the collective traffic response.

Together, these results establish the effective temperature as a physics-based indicator of urban traffic resilience: a single, dimensionless, city-level parameter derived from first principles that orders cities by their vulnerability to congestion regime shifts. Unlike empirical congestion indices, which measure average outcomes, $T$ characterizes the structural susceptibility of a city to demand-driven transitions, making it actionable for infrastructure planning, mobility policy, and the design of demand management interventions.

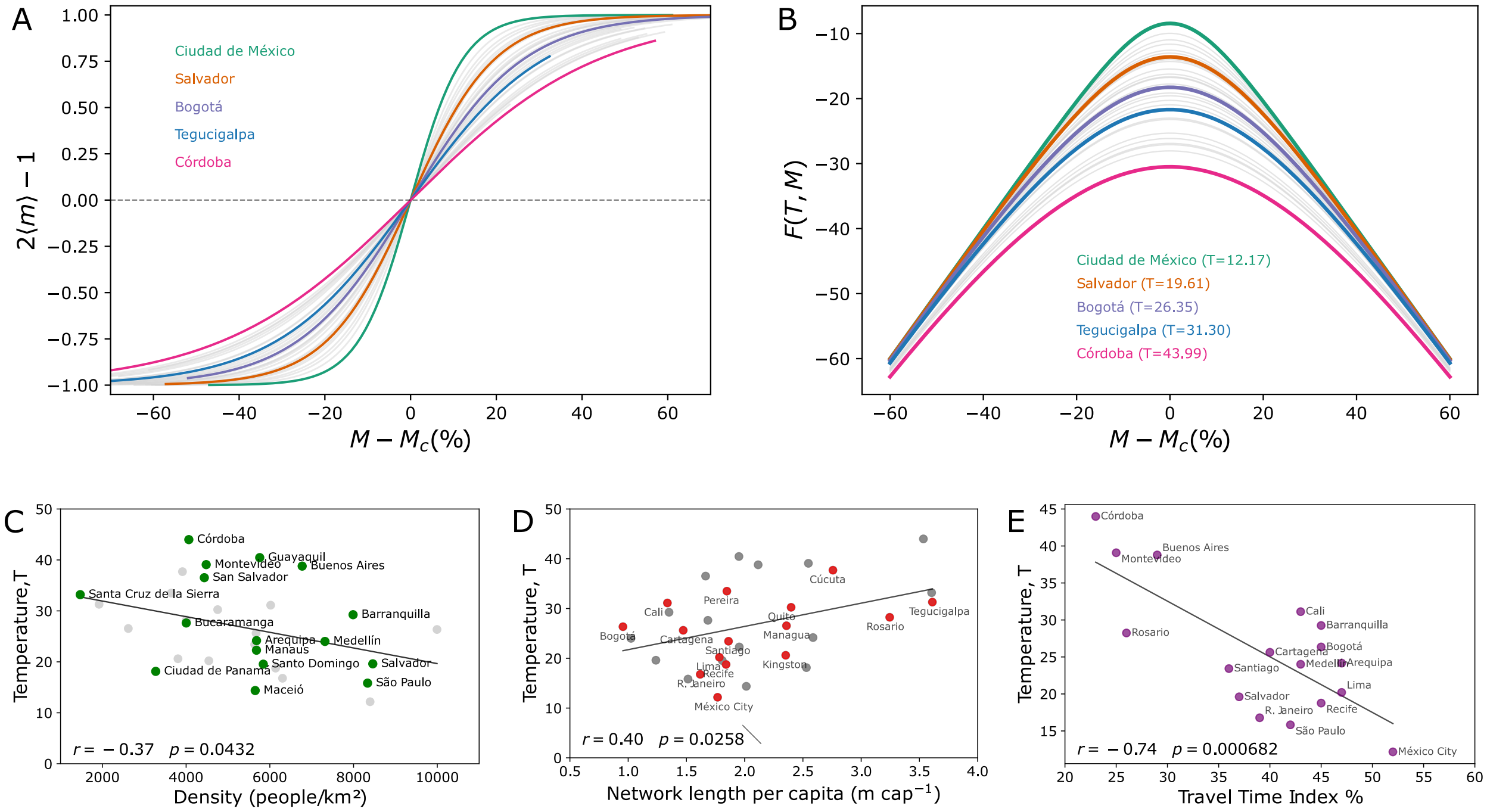

**Figure 3. Effective temperature as a structural indicator of urban traffic resilience.** (A) Congestion responses for multiple cities, expressed as the normalized order parameter $2\langle m\rangle - 1$ as a function of mobility relative to the city-specific transition point, $M - M_c$. The collapse of diverse cities onto a common functional form — differing only in transition width — demonstrates that a single parameter, the effective temperature $T$, captures systematic cross-city differences in congestion response. (B) Reconstructed effective free-energy profiles $F(T, M)$ associated with the canonical description. The free energy displays a smooth maximum near $M_c$ and decreases toward saturated congestion regimes, consistent with a field-driven organization of urban traffic states; narrower profiles correspond to lower-temperature, more congestion-fragile cities. (C) Effective temperature as a function of population density. Denser cities exhibit lower effective temperatures, corresponding to sharper mobility–congestion transitions and reduced configurational flexibility. (D) Effective temperature as a function of road network length per capita. Cities with greater infrastructure availability per person display higher effective temperatures and more gradual congestion responses, reflecting a broader exploration of network configurations under increasing demand. (E) Effective temperature versus the annual congestion index reported by TomTom Traffic Index 2023 (19). Lower effective temperatures are associated with systematically higher chronic congestion levels, directly linking the thermodynamic parameter to observable urban outcomes and confirming its interpretation as an indicator of congestion resilience. Solid lines indicate linear fits; reported r and p values denote Pearson correlation statistics.

## Free energy, capacity, and the statistical origin of congestion

The Helmholtz free energy $F = \langle E\rangle - TS$ characterizes the balance between energetic bias and entropic dispersion in systems with many accessible configurations. he macroscopic state is determined by minimizing $F$ with respect to internal degrees of freedom at fixed external constraints; the free-energy profile as a function of the external field encodes how the configurational landscape evolves as forcing increases. We argue that urban traffic networks admit a description of precisely this form.

The natural effective energy variable is the volume-over-capacity ratio ($VOC$). Unlike raw traffic volume, $VOC$ normalizes local flow by local infrastructure capacity, collapsing structural heterogeneity across road types onto a common dimensionless scale: a link with $VOC \to 0$ is near its ground state; $VOC \to 1$ operates at capacity; $VOC > 1$ is overloaded. The network-level mean $\langle VOC \rangle$ therefore plays the role of effective internal energy — measuring how far the collective traffic state departs from uncongested conditions. This identification is not merely formal: previous simulation studies have shown that the distribution of $VOC$ across links is approximately exponential, whereas the distribution of raw traffic volume is not (24,25) — a pattern that does not arise trivially from routing, but from the interaction between heterogeneous capacities, fixed aggregate demand, and congestion-dependent travel costs.

The exponential form has a natural statistical origin. In user-equilibrium assignment, total demand is fixed but the microscopic allocation of flows across routes is not unique — multiple link-level configurations satisfy the same equilibrium conditions, generating a large ensemble of admissible microstates (8,26,27). Entropy maximization under a constraint on mean demand leads naturally to an exponential distribution $P(VOC) \propto e^{-VOC/T}$ (Boltzmann-like form), where $T$ captures variability in demand and routing. Critically, this ensemble is not hypothetical: urban traffic is recurrent. Day after day, the same network absorbs similar aggregate demand with variations in origin-destination patterns, departure times, and route choices that effectively sample the configuration space. Drivers' accumulated routing experience reinforces this sampling (21,22), so the daily traffic state is best understood as a draw from a canonical ensemble — and $T$ as a structural property of the city estimated from regularities across many daily realizations.

Effective free-energy profiles reconstructed from mobility–congestion data display structural features reminiscent of the macroscopic fundamental diagram (MFD) (2–6): nonlinear response, characteristic capacity-like points, and saturation under increasing demand. Within the thermodynamic framework, this resemblance has a precise interpretation — the MFD is not a primitive empirical law but a macroscopic response function emerging from the statistical organization of congestion states. Specifically, the MFD corresponds to the envelope of $\langle VOC \rangle$ over the free-energy landscape: a projection that captures throughput but discards the configurational entropy encoded in the full description. The appropriate macroscopic object is therefore not flow versus density, but the effective Helmholtz free energy — which jointly encodes both.

**Empirical test of the thermodynamic framework using flow–density data**

To evaluate the thermodynamic framework directly using independent data, we analyzed high-resolution flow and density measurements from fixed loop detectors across eight large metropolitan areas: Bordeaux, London, Los Angeles, Madrid, Paris, Rotterdam, Taipei, and Torino (28). These datasets have previously supported empirical studies of the macroscopic fundamental diagram (3,4,29) and provide a test of the framework under normal traffic conditions — entirely independent of the pandemic-period mobility data used in the city-scale analysis.

We begin by testing whether the sigmoidal congestion transition observed at the city scale is reproduced at the link level. Within each 15-minute window, each detector lane is classified as congested when its density exceeds a road-type-specific threshold— reflecting the empirically observed onset of speed breakdown for each functional class (see Methods and Supplementary Note 2). The fraction of congested lanes $p_{jam}$ and mean network density $\rho$ are then computed across all monitored links. Across all eight cities, $p_{jam}$ exhibits a sigmoidal dependence on $\rho$ (Fig. 4A shows Bordeaux as a representative example; full panel in Supplementary Figure S7):

$$p_{jam}(\rho) = \frac{p_{max}^{(emp)}}{2}\left[1 + \tanh\left(\frac{\rho - \rho_c^{(emp)}}{\Delta}\right)\right]. \qquad (4)$$

Here, $\rho_c^{(emp)}$ is the characteristic transition density — the network density at which half the monitored links are congested — estimated directly from the sigmoidal fit to the data. Δ is the transition width controlling the sharpness of the transition, and $p_{max}^{(emp)}$ is the empirically observed saturation level of the jam fraction, reflecting the maximum fraction of links that become congested under high demand conditions. Both $\rho_c^{(emp)}$ and $p_{max}^{(emp)}$ are city-specific parameters estimated independently for each metropolitan area (Supplementary Table S1).

The consistency of this functional form across two independent datasets — floating-car data during the pandemic and loop detectors under normal conditions, spanning three continents — establishes the sigmoidal congestion transition as a robust macroscopic property of urban traffic rather than an artifact of any specific measurement approach or historical period.

We next test whether $VOC$ follows the Boltzmann-like statistics anticipated by the thermodynamic framework. For each city and each hour of the day, we compute the distribution of normalized $VOC$, $x = \frac{VOC}{\langle VOC \rangle_t}$, pooling observations across days. Despite strong hourly changes in traffic demand, the functional form remains stable across all 24 hours (Fig. 4B, gray curves) and is well described by

$$P_t(\mathrm{VOC}) \propto exp\left[-\alpha_t \frac{\langle \mathrm{VOC} \rangle_t}{VOC} - \beta_t \frac{VOC}{\langle \mathrm{VOC} \rangle_t}\right], \qquad (5)$$

At large $x$, this reduces asymptotically to $P_t(\mathrm{x}) \sim exp\,[-\beta_t x]$ — a Boltzmann form with effective temperature, $T_{VOC,t} = \frac{1}{\beta_t}$. The parameter $\beta_t$ remains statistically stable across working hours (Fig. 4B inset, blue points), confirming that $T_{VOC}$ is a structural property of the city rather than a transient feature of specific demand levels. The parameter $\alpha_t$, which suppresses very low normalized flows, reflects detector placement bias toward congestion-prone corridors rather than a fundamental thermodynamic property (Supplementary Note 2).

The thermodynamic framework is validated by the agreement between two effective temperatures derived from entirely different observables. $T_{VOC}$, extracted from the exponential decay of link-level VOC distributions, and $T_{jam} = \frac{\Delta}{\rho_c^{(emp)}}$, inferred from the width of the macroscopic density-jam transition, agree closely across all eight cities (Fig. 4C) — with deviations only where the density transition is partially observed within the available range. This convergence — between a link-level energy statistic and a macroscopic order-parameter transition — provides direct evidence that the thermodynamic description captures a genuine structural property of urban traffic, not a fitting artifact.

Having identified the order parameter $p_{jam}$, and the effective energy variable ($VOC$), we reconstruct the free-energy landscape in coarse-grained form as

$$F(\Delta, \rho) \sim \kappa_{city} \left( \langle VOC \rangle - \frac{\Delta}{\rho_{th}} S \right) \qquad (6)$$

where $S(p_{jam}) = -p_{jam}\, ln\, p_{jam} - (1 - p_{jam}) ln(1 - p_{jam})$ is the Gibbs entropy of the binary congestion state — maximal when congested and uncongested links are approximately balanced, and minimal when congestion is either highly localized or nearly saturated. The fator $\kappa_{city}$ absorbs sampling effects arising from detector placement, which tends to overrepresent congestion-prone corridors of the road network. As derived in Supplementary Note 4, this free-energy form follows from three minimal ingredients: normalized network throughput scaling with density, mean velocity decreasing approximately linearly with jam fraction, and the binary classification of congestion states — with no assumptions about microscopic origin-destination patterns.

The quantity $\rho_{th}$ denotes the average jam-density threshold across monitored road segments, computed from road-type-specific thresholds (Methods, Supplementary Note 2 ). It reflects structural properties of the network rather than instantaneous traffic conditions, providing the characteristic density scale that links the energetic and entropic contributions in Eq. 6. With full network coverage, $\rho_c^{(emp)}$ would be expected to approach $\rho_{th}$— in that limit, normalizing the transition width by $\rho_c^{(emp)}$ naturally absorbs this structural scale, leaving $T_{jam} = \frac{\Delta}{\rho_c^{(emp)}}$ as the dimensionless parameter that captures congestion dynamics across cities.

Fig. 4D shows the reconstructed $F(\rho)$ computed directly from Eq. 6 using empirical $\langle VOC \rangle$ and $S(p_{jam})$, alongside the canonical fitting form

$$F(\Delta, \rho) = F_0 - \frac{p_{max}^{(fit)}}{2} \Delta \ \ln \left[ 2 \cosh \left( \frac{\rho - \rho_c^{(fit)}}{\Delta} \right) \right]. \qquad (7)$$

Here $F_0$ is an additive offset, $\rho_c^{(fit)}$ is the fitted critical density, $p_{max}^{(fit)}$ is the fitted saturation level and $\Delta$ is fixed from the empirical transition width estimated in Eq. 4 — it is not a free parameter. The close agreement between the two curves — one constructed directly from data via Eq. 6, the other from the canonical form of Eq. 7 — confirms that the thermodynamic description is not imposed on the data but emerges from it.

The right axis of Fig. 4D shows $\langle VOC \rangle$ as a function of density — the network-level analog of the macroscopic fundamental diagram. The free energy reaches its maximum at a density systematically lower than the peak in $\langle VOC \rangle$ revealing that large-scale configurational reorganization of the network precedes the classical capacity drop. Structural instability develops before it becomes visible in conventional throughput-based measures.

Taking the derivative $-\frac{\partial F}{\partial \rho}$, recovers the sigmoidal equation of state, reproducing the observed variation of $p_{jam}$ (Fig. 4E). The fitted critical density $\rho_c^{(fit)}$ agrees closely with the independently extracted empirical value $\rho_c^{(emp)}$ (Fig. 4F), providing an internal consistency check of the framework. Together, Figs. 4D–F demonstrate that a single free-energy function simultaneously encodes the macroscopic congestion transition, the link-level energy statistics, and the network throughput curve — unifying descriptions that were previously treated as separate empirical relations.

Within this framework, the macroscopic fundamental diagram emerges as the envelope of $\langle VOC \rangle$ over the free-energy landscape: a projection that captures throughput but discards the configurational entropy encoded in the full description. The free energy is the more complete macroscopic object

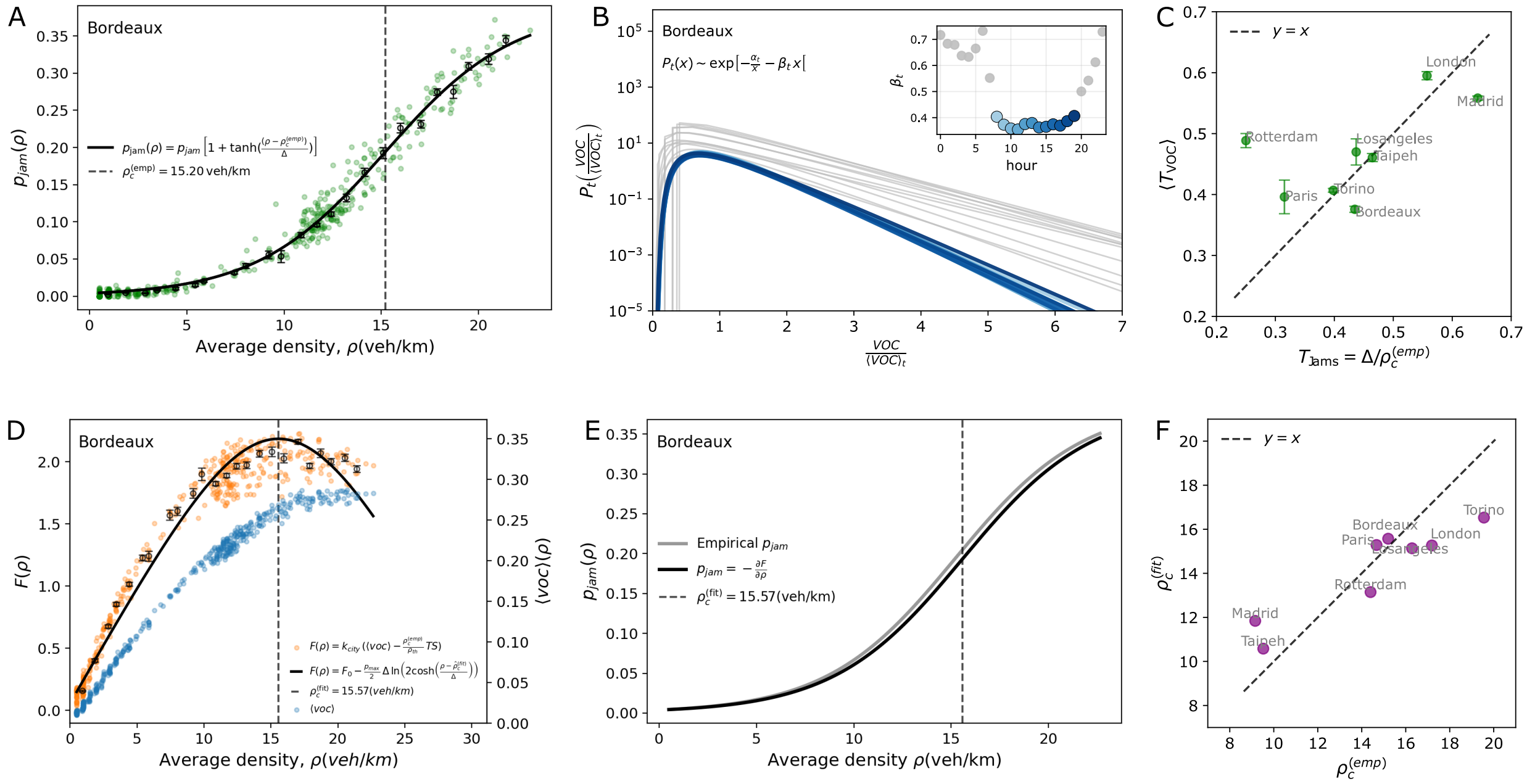

**Figure 4. Empirical reconstruction of the thermodynamic congestion** framework (**as representative example; full panel in SI Appendix, Fig. S7).** (A) Fraction of jammed links as a function of mean network density. Points correspond to aggregate 15-minute observations, while the solid curve shows the sigmoidal fit used to extract the critical density $\rho_c^{(emp)}$ and transition width Δ. (B) Empirical distribution of normalized volume-over-capacity, $x = \frac{VOC}{\langle \mathrm{VOC} \rangle_t}$, across hours of the day. For each hour t, the distributions are well described by $P(\mathrm{x}) \propto exp[-\alpha_t/x - \beta_t x]$. Inset: hourly estimates of the decay parameter $\beta_t$. The stability of $\beta$ during working hours (blue points) confirms that the effective temperature $T_{VOC} = 1/\beta$ is a structural property of the city rather than a transient feature of specific demand levels. (C) Comparison between effective temperature estimate obtained from independent observables: $T_{VOC}$, extracted from the exponential decay of VOC distributions, and $T_{jam} = \frac{\Delta}{\rho_c^{(emp)}}$ inferred from the width of the macroscopic congestion transition. The close agreement along the identity line (dashed) confirms that link-level energy statistics and the macroscopic order-parameter transition are governed by the same structural parameter. (D) Reconstructed effective free energy F(ρ) — computed directly from empirical $\langle VOC \rangle$ and binary entropy $S(p_{jam})$ via Eq. 6 (points) and canonical fitting form Eq. 7 (black curve) — alongside mean $\langle VOC \rangle$ as a function of network density (right axis, blue points), the network-level analog of the macroscopic fundamental diagram. The free energy reaches its maximum at a density lower than the peak in $\langle VOC \rangle$ revealing that configurational reorganization of the network precedes throughput collapse. (E) Empirical jam fraction $p_{jam}$ (gray) compared with −∂F/∂ρ derived from the reconstructed free energy (black), demonstrating that the thermodynamic potential reproduces the observed sigmoidal congestion transition. (F) Critical densities inferred independently from the free-energy reconstruction $\rho_c^{(fit)}$ and from the empirical jam transition $\rho_c^{(emp)}$ across all eight cities. The close agreement along the identity line (dashed) provides an internal consistency check of the framework.

## DISCUSSION

Cities do not simply accumulate congestion as demand increases — they reorganize. Our results show that urban traffic behaves as a collective system whose macroscopic states are shaped by the interplay between capacity constraints and the redistribution of flows across the network, rather than by the independent failure of isolated bottlenecks. Because urban networks are repeatedly loaded by daily mobility demand, they effectively sample an ensemble of traffic configurations, enabling an equilibrium-like statistical description at the macroscopic scale. This provides a concrete realization of the program called for by Prigogine and Herman nearly five decades ago (1): a statistical-mechanical description linking the collective behavior of traffic to stable macroscopic regularities.

A central result is the separation between network-level reorganization and throughput degradation. The thermodynamic transition in jam fraction precedes the peak in average volume-over-capacity, indicating that large-scale configurational reorganization develops before conventional aggregate metrics register a clear capacity loss. This ordering has a practical consequence: free-energy-based observables provide early indicators of systemic fragility that are invisible to standard throughput-based measures. A city approaching its thermodynamic transition may appear operationally normal by conventional metrics while already being structurally vulnerable to abrupt congestion escalation.

The effective temperature T introduced here quantifies this structural vulnerability. Cities with higher T exhibit broader congestion transitions and greater configurational flexibility — they absorb demand gradually and distribute congestion across a wider range of network states. Low-temperature cities display sharper transitions and earlier saturation, concentrating congestion on a limited subset of structurally critical corridors. Crucially, configurational flexibility is distinct from raw capacity: a city may sustain substantial traffic volumes while remaining fragile in its collective organization. The correlations between T, infrastructure supply, population density, and chronic congestion levels establish it as a compact, physics-based indicator of urban traffic resilience — complementing existing congestion indices with a measure of structural susceptibility rather than average outcome.

The sigmoidal mobility–congestion relation has direct implications for demand management and policy. The transition identifies a city-specific threshold beyond which congestion grows rapidly toward saturation — and the position of a city along this curve quantifies how much additional demand can be absorbed before escalation becomes sharp. Because the response is strongly nonlinear, modest reductions in vehicle activity produce limited gains when the system operates on the saturated branch. This may explain why many traffic-reduction measures yield weaker-than-expected outcomes: unless demand falls sufficiently to move the system back across the transition, the collective network state remains in the congested regime. Conversely, interventions targeted at increasing T — through infrastructure investment, network redundancy, or demand redistribution — may shift the transition threshold and expand the range of sustainable mobility.

Several limitations of the present framework point toward natural extensions. The mean-field-like description treats road segments as statistically independent, neglecting explicit interactions such as spillbacks and queue propagation. Allowing explicit inter-segment couplings would modify the effective potential and could account for hysteresis, metastability, and spatially clustered congestion patterns observed in specific cities. The transition in jam fraction may also be related to percolation-like phenomena, in which localized disruptions coalesce into network-spanning congested structures (11–13) — a connection that remains to be formalized within the thermodynamic framework. The role of origin–destination structure is another open question: microscopic OD heterogeneity may enter implicitly through the entropy of congestion configurations, but how changes in OD patterns reshape this entropic contribution remains to be determined.

Congestion, in this view, is not merely a failure of traffic but the structural cost of coordinating recurrent mobility at scale. Urban road networks organize into macroscopic states shaped by entropy and capacity constraints, from which classical relations — including the macroscopic fundamental diagram — emerge as aggregate response functions rather than primitive laws. The fundamental diagram is not wrong; it is incomplete. The appropriate macroscopic description is the effective free energy, which encodes both the throughput captured by the diagram and the configurational organization that the diagram discards. Identifying this structure opens a path toward physics-based indicators of urban resilience, principled comparisons across cities, and a quantitative foundation for anticipating how cities will respond to shifts in mobility demand.

## Methods

**City-scale mobility, congestion, and day selection.** City-scale congestion was measured using the Traffic Congestion Index (TCI) (17), derived from aggregate Waze jam-line data reported at 5-minute resolution and released by the Inter-American Development Bank (IDB) (30). The dataset covers 46 major cities across Latin America and the Caribbean during the COVID-19 period — a natural experiment in which mobility demand experienced abrupt, large-amplitude fluctuations that allowed urban systems to traverse traffic regimes rarely accessible under normal conditions. At each 5-minute snapshot, Waze identifies contiguous road segments operating below free-flow speed; their total length is accumulated over space and time to produce a daily indicator capturing both the spatial extent and temporal persistence of congestion. Daily TCI was expressed as a percentage change relative to a pre-pandemic baseline using day-of-week matching, following the IDB definition.

Mobility demand was measured using the Google Community Mobility Reports (18) workplace indicator — daily percentage changes in work-related activity relative to a pre-pandemic baseline. The workplace category was selected because commuting accounts for a substantial and structurally stable share of daily urban travel, making it a reliable proxy for aggregate mobility demand that is less sensitive to discretionary trip variation than other categories. The mobility and congestion series were aligned at daily resolution over their common observation window. Because both variables are expressed as relative percentage changes, moderate differences in polygon boundaries between providers are not expected to materially affect the city-level relationships analyzed here.

TomTom travel-time indices (19,20) and Waze vehicle-miles-traveled (VMT) series (20) were used only as external benchmarks and were not used to define the main congestion metric. To isolate recurrent commuting-driven traffic dynamics from exceptional disruptions, weekends and public holidays were excluded — weekend traffic patterns reflect discretionary rather than commuting demand and would introduce systematic heterogeneity into the mobility–congestion relationship. Cities exhibiting prolonged anomalous periods during the study window were considered separately when assessing the quality of the sigmoidal fit. All city-scale analyses relied exclusively on provider-processed aggregate indicators; no raw jam geometries, individual trajectories, or user-level records were accessed.

**Detector-based traffic data.** Link-scale traffic states were analyzed using the UTD19 dataset (28), compiled by the Institute for Transport Planning and Systems (IVT) at ETH Zurich. UTD19 contains stationary loop-detector measurements from urban roads in more than 40 cities worldwide, including vehicle flow, occupancy, speed, and link-level metadata. For the cities analyzed here, detector records correspond to pre-COVID years, providing a test of the thermodynamic framework under normal demand conditions entirely independent of the pandemic-period city-scale analysis. Most cities are reported at 3–5 minute aggregation intervals; Paris is available at hourly resolution. From the full dataset, we retained eight metropolitan areas — Bordeaux, London, Los Angeles, Madrid, Paris, Rotterdam, Taipei, and Torino — selected on the basis of detector network density sufficient to characterize collective network-level traffic states rather than isolated corridor behavior, and data completeness across multiple years.

**Capacity assignment, jam classification, and network observables.** Lane-level density was estimated from loop-detector occupancy — defined as the fraction of time a detector was occupied within each aggregation interval — and converted into vehicles per unit length using the effective detector segment length and an assumed average vehicle length, yielding estimates in veh km$^{-1}$ lane$^{-1}$. Network density was computed as the median across all available detector lanes at each time interval. The median was preferred over the mean because the empirical distribution of lane densities is heavy-tailed: a small number of highly congested links can disproportionately inflate the mean, whereas the median provides a more robust estimate of the typical macroscopic load of the network.

Volume over capacity was defined as $VOC = \frac{q}{C}$, where $q$ is the observed lane-level flow and $C$ is the nominal lane-level capacity assigned to each road segment according to its OpenStreetMap (31) functional road classification. Network-level $VOC$ was computed as the mean across detector lanes at each time interval. Jam fraction was defined as the fraction of monitored detector lanes classified as congested at a given time interval. A detector lane was classified as congested when its density exceeded a road-type-specific threshold $\rho_{th}$, reflecting the empirically observed onset of speed breakdown for each functional class. Jam fraction $p_{jam}$ was defined as the fraction of monitored lanes classified as congested within each time window. Both capacity values and congestion thresholds were assigned by functional class using representative values for typical urban conditions; full mappings are reported in Supplementary Note 2. Main results were robust to reasonable variations in these assumptions.

**Mobility–congestion fits and canonical representation.** Macroscopic relationships were estimated using fixed-width binning in the relevant control variable, followed by nonlinear least-squares fitting using weighted least squares with bin-level standard errors as fit weights — down-weighting poorly sampled bins without excluding them. In the city-scale analysis, the empirical TCI–mobility relation was fitted with a sigmoidal function to estimate the characteristic mobility $M_c$, transition width, and saturation level. For comparison with the canonical two-state formulation, fitted congestion curves were normalized by their saturation values and transformed into the centered order parameter $2\langle m \rangle - 1$, mapping congestion onto the symmetric interval $[-1,1]$ consistent with the binary classification of road segments. For full details of the estimation procedure see Supplementary Note 3.

## Acknowledgments

The author thank the Inter-American Development Bank (IDB) for providing access to the Traffic Congestion Index dataset used in this study. We also acknowledge the authors of the UTD19 traffic detector dataset for making large-scale urban traffic measurements publicly available, which were used for complementary analyses in this work.

**Funding:** This research received no external funding.

**Competing interests:** The author declare that they have no competing interests.

**Data, code, and materials availability:** Traffic congestion data are derived from the Inter-American Development Bank Traffic Congestion Index dataset based on Waze for Cities Data. Traffic detector measurements are obtained from the publicly available UTD19 dataset. Processed datasets and analysis code supporting the findings of this study will be deposited in a permanent public repository upon publication.

Supplementary Information for

# Title: Thermodynamic phase transitions reveal the resilience structure of urban traffic congestion

## Contents

# Supplementary Notes

## Supplementary Note 1: Data Source

Urban traffic conditions were characterized using a combination of floating-car data and fixed traffic detector measurements. The analysis relies exclusively on provider-processed and publicly released aggregate indicators; no raw data reconstruction or individual-level processing was performed. Because the spatial units used in the COVID-period mobility datasets do not necessarily coincide exactly with the metropolitan polygons defined in the congestion datasets, perfect spatial alignment cannot be verified. However, since the mobility indicators are expressed as relative percentage changes rather than extensive quantities, moderate variations in polygon boundaries are not expected to materially affect the congestion–mobility relationships analyzed here.

*Traffic Congestion Index (TCI)*

Traffic congestion is measured using the Traffic Congestion Intensity (TCI) as defined in the Inter-American Development Bank (IDB) Coronavirus Impact Dashboard framework (1). The IDB accesses aggregate Waze data through the Waze for Cities Program.

Waze identifies traffic jams as contiguous road segments ("jam lines") where observed speeds fall below expected free-flow conditions. Jam lines are reported at 5-minute intervals. For each metropolitan area (polygon p), total jam length at interval i is defined as

$$JAM_{ip} = \sum_{j} L_{jip}, \qquad (1)$$

where $L_{jip}$ is the length of jam line j within polygon p.

Daily Traffic Congestion Intensity is computed as:

$$TCI_{pt} = \sum_{i \in t} JAM_{ip} \qquad (2)$$

where t denotes a full day. By construction, $TCI_{pt}$ captures both the spatial extent and temporal persistence of congestion (see Supplementary Fig. S1 for a schematic illustration).

To evaluate changes over time, the IDB defines March 2–8, 2020 as the baseline week and applies day-of-the-week matching. The percentage change in congestion is defined as

$$\Delta TCI_{pdt1} = \left(\frac{TCI_{pdt1}}{TCI_{pdt0}} - 1\right) \times 100. \qquad (3)$$

This study relies exclusively on aggregate $\Delta TCI_{pdt1}$ indicators as reported by the IDB. For simplicity, in the main text we denote this quantity as $TCI$. No raw jam geometries or user-level data are accessed.

The dataset metadata include the polygon boundaries, total road network length, metropolitan population, $TCI_{pdt1}$, and $TCI_{pdt0}$. Selected descriptive properties of these variables are shown in Supplementary Fig. S1.

*TomTom Traffic Index and Vehicle Miles Traveled (VMT)*

The TomTom Traffic Index (2) is constructed from proprietary floating-car data collected through navigation devices, connected vehicles, and mobile applications at the city scale. The index measures the additional travel time experienced by drivers relative to free-flow conditions due to congestion and is reported as a percentage increase over free-flow travel time.

Waze also provides a city-level measure of Vehicle Miles Traveled (VMT), representing total driving activity aggregated across users. VMT is expressed as relative changes with respect to predefined baseline periods.

Both indicators are expressed relative to free-flow conditions and are used in this study solely as external benchmarks for comparative analysis. They are not used to compute the main congestion metric introduced here (see Fig. 2B–C in the main text and Supplementary Figs. S5–S6).

Historical city-level time series for both metrics were obtained from archived mobility repositories compiled during the COVID-19 period (3).

In addition, the TomTom Traffic Index value shown in Fig. 2E corresponds to the official 2023 city congestion ranking (2), as published on the TomTom Traffic Index website. This value is used as a cross-sectional reference and is not part of the time-series analysis.

*Google COVID-19 Community Mobility Reports*

Mobility data were obtained from the Google COVID-19 Community Mobility Reports(4)**,** which provide daily percentage changes in visits to different place categories relative to a pre-pandemic baseline. The data are generated from aggregated, anonymized location information from users who have opted into Google Location History.

Here, only the *Workplaces* category is used. This indicator reflects daily relative changes in work-related activity and commuting demand and is treated as the mobility variable. The *Workplaces* time series were restricted to the common observation window (March 9, 2020 to May 5, 2022) and aligned with congestion indicators at daily resolution. When required for comparative analysis, mobility was linearly rescaled to a dimensionless form without altering its temporal structure.

Throughout the analysis, the Workplaces indicator is interpreted as a control parameter (M in main text) governing the macroscopic state of the urban traffic system, while congestion metrics quantify the system's aggregate response.

*Urban traffic detector dataset* (UTD19)

Fixed traffic detector data were obtained from the UTD19 dataset (5)**,** a large-scale urban traffic dataset compiled by the Institute for Transport Planning and Systems (IVT) at ETH

Zurich**.** UTD19 consists of measurements from stationary detectors (loop detectors) deployed on urban roads in more than 40 cities worldwide. The dataset includes traffic variables such as vehicle flow, occupancy, and speed, reported at short aggregation intervals (typically 3–5 minutes; 1 hour for Paris). It also provides link-level metadata, including segment length, number of lanes, and road functional classification. The dataset spans multiple years and comprises billions of detector observations.

These data have been widely used to study empirical properties of urban traffic, including the macroscopic fundamental diagram (6–9). In this study, no additional processing was applied beyond temporal aggregation consistent with other data sources and the conversion of occupancy measurements into density estimates. From the full set of cities, we retained those with sufficiently dense detector coverage across the urban area. This filtering resulted in eight cities: Bordeaux, London, Los Angeles, Madrid, Paris, Rotterdam, Taipei, and Torino.

**Supplementary Note 2**: **Definition of macroscopic variables**

Here we define the macroscopic observables used to characterize the collective state of urban traffic systems.

*Density ρ*

Road segment density ρ(veh/km) is estimated from loop-detector occupancy measurements. Occupancy O, is defined as the fraction of time during which a given lane detector is occupied within an aggregation interval. For each detector lane, density is computed as

$$\rho = \frac{O}{Ldet+Lv}\, x1000 \quad (4)$$

where:

- O is occupancy in fractional form,
- $Ldet$ is the effective detector segment length (m),
- $Lv$ is the assumed average vehicle length (m),
- the factor 1000 converts units from veh/m to veh/km.

Because loop detectors operate at the lane level, density is defined in units of veh/km/lane. Network-level density is obtained by aggregating lane-level densities using the median across all available detector lanes at each time interval. The median is used instead of the mean because the empirical distribution of densities exhibits a heavy-tailed (approximately power-law) behavior. Thus, the mean can be disproportionately influenced by extreme values, whereas the median provides a more robust estimate of the typical macroscopic load of the network.

*Volume-over-Capacity (VOC)*

Volume-over-capacity is defined as $VOC = \frac{q}{C}$, where q is observed flow and C is the assumed lane-level capacity of the corresponding road segment.

Capacity values are assigned according to the OpenStreetMap functional road classification of each link. A fixed capacity map is used:

- motorway: 2200 veh/h/lane
- trunk: 2400
- primary: 2200
- secondary: 1900
- tertiary: 1400
- living_street: 400
- unclassified: 700
- motorway_link: 2000
- trunk_link: 2000
- primary_link: 1800
- secondary_link: 1500
- residential: 900
- service: 600
- other: 1000

Capacities are expressed in vehicles per hour per lane and are treated as nominal operating limits representative of typical urban conditions. Network-level VOC is computed as the mean across detector lanes at each time interval. The main results are robust to moderate variations in the assumed capacity values, as such variations produce only small rescalings of VOC without altering the observed macroscopic relationships.

*Jam Fraction* $p_{jam}$

The jam fraction $p_{jam}$ is defined as the fraction of monitored links classified as congested at a given time interval, $p_{jam} = \frac{N_{jam}}{N_{total}}$, where $N_{jam}$ is the number of detector lanes exceeding a congestion threshold and $N_{total}$ is the total number of monitored lanes. A lane is classified as congested when its density exceeds a road-type–specific threshold, $\rho > \rho_{th}$.

The density thresholds $\rho_{th}$ (veh/km/lane) are assigned according to functional road classification, consistent with critical density ranges reported in the traffic flow literature for freeways, urban arterials, and local streets:

- motorway, motorway_link, trunk, trunk_link: 45 veh/km/lane
- primary: 45
- primary_link: 35
- secondary: 33
- secondary_link: 30
- tertiary, residential, living_street, service, unclassified, other: 30

These values fall within empirically observed ranges of critical density associated with the onset of congestion across road classes. Results are insensitive to reasonable variations in the assumed density thresholds values.

**Supplementary Note 3: Statistical Estimation Procedures**

This section describes the procedures used to estimate the macroscopic transition and associated parameters.

*Binning Protocol*

All macroscopic relationships were estimated using fixed-width binning in the corresponding control variable (e.g., density ρ or mobility M). For each city, observations were grouped into equal-width bins, and the mean value of the associated macroscopic observable (e.g., jam fraction $p_{jam}$ percentage change in TCI, or reconstructed free-energy derivatives) was computed within each bin. Bins with insufficient observations were excluded. This coarse-graining procedure was applied consistently across all sigmoidal and free-energy analyses.

*Estimation of the Sigmoidal Transition*

Sigmoidal transitions were estimated by nonlinear least-squares fits using `curve_fit`.

For the jam fraction–density relationship, we used a hyperbolic tangent form:

$$p_{jam}(\rho) = \frac{p_{max}}{2}\left[1 + tanh\left(\frac{\rho-\rho_c^{(emp)}}{\Delta}\right)\right], \qquad (5)$$

where $\rho_c^{(emp)}$ is the transition midpoint and $\Delta$ controls the transition width.

An analogous procedure was applied to the congestion–mobility relationship using the Traffic Congestion Intensity (TCI). The empirical TCI–mobility curve was fitted with the same functional form, with L corresponding to the saturation value TCImax.

To enable comparison with the canonical two-state formulation, the fitted TCI was normalized as $\langle m\rangle = \frac{TCI}{TCI_{max}}$, and transformed into a centered order parameter, 2⟨m⟩−1.

Fits were performed using `curve_fit` with weighted least squares, providing bin-level uncertainties σ (std error) as weights (via the `sigma` argument). Goodness-of-fit ($R^2$) values were computed using the binned median data used for parameter estimation.

**Supplementary Note 4: Free-energy derivation of the sigmoidal jam transition**

*Effective free-energy formulation*

We describe the macroscopic congestion state of the monitored road network through the fraction of jammed links ppp, treated as a collective order parameter. The mean network density ρ acts as the externally controlled variable.

We introduce an effective free energy $F(\Delta,\rho) \sim \kappa_{city}\left(\langle VOC\rangle - \frac{\Delta}{\rho_{th}} S\right)$ (Eq. 5 in the main text), where $S(p) = -p_{jam}\, ln\, p_{jam} - (1-p_{jam}) ln(1-p_{jam})$ is the binary entropy associated with jammed and free links. Here Δ represents an effective temperature (expressed in density units) controlling the width of the transition, and $\rho_{th}$ is the characteristic jam-density threshold defining the binary congestion state of a road segment.

In the main text we define the transition width parameter $\Delta \equiv T_{jam}\rho_c^{(emp)}$.

*Energetic term and throughput interpretation*

The energetic contribution is taken proportional to the average $\langle E\rangle = \langle VOC\rangle = \langle \frac{q}{C}\rangle$, where $q$ denotes traffic flow and C the link capacity.

Using the macroscopic traffic identity $\langle q\rangle = \rho\, v(\rho, p_{jam})$, and approximating $\langle \frac{q}{C}\rangle \approx \frac{\langle q\rangle}{\langle C\rangle}$, we obtain

$$\langle E\rangle(\rho, p_{jam}) \propto \frac{\rho\, v(\rho, p_{jam})}{\langle C\rangle}. \quad (6)$$

Consistent with classical two-fluid descriptions of traffic, we empirically observe an approximately linear relation between average velocity and the fraction of jammed links (Supplementary Fig. S8),

$$v(\rho, p_{jam}) = v_{max}[1 - \gamma p_{jam}] \qquad (7)$$

where γ quantifies the sensitivity of average velocity to the fraction of jammed links. This equation resembles the two-fluid formulation of urban traffic proposed by Prigogine and Herman(10), where the macroscopic speed of the system is determined by the fraction of vehicles that belong to the congested phase.

At the link level, the maximum observed flow $q_{\ max}$ is the observed corresponds to the operational capacity of the link. Under a triangular fundamental-diagram approximation, the characteristic maximum flow satisfies

$$\langle q_{\ max}\rangle \approx \rho_{th}\, v_{max} \quad (8)$$

Because designed link capacities are prescribed by road class, this scale is proportional to the nominal capacity, so that

$$\rho_{th}\, v_{max} \approx \eta\langle C\rangle \qquad (9)$$

where η= O(1) is a dimensionless proportionality factor that absorbs deviations from an ideal triangular diagram and the mapping between the threshold density imposed in our classification and the density at capacity.

Substituting this relation yields

$$\langle E\rangle(\rho, p_{jam}) = \frac{\rho}{\eta\,\rho_{th}}(1-\gamma p_{jam}). \quad (10)$$

Hence

$$\left.\frac{\partial\langle E\rangle}{\partial p_{jam}}\right|_{\rho} = -\frac{\gamma}{\eta}\frac{\rho}{\rho_{th}}. \qquad (11)$$

*Minimization and logistic form*

Equilibrium at fixed density ρ is obtained from

$$\left.\frac{\partial F}{\partial p_{jam}}\right|_{\rho} = 0, \quad (12)$$

which gives

$$\left.\frac{\partial\langle E\rangle}{\partial p_{jam}}\right|_{\rho} - \frac{\Delta}{\rho_{th}}\frac{\partial S}{\partial p_{jam}} = 0. \qquad (13)$$

Since

$$\frac{\partial S}{\partial p_{jam}} = -\ln\frac{p_{jam}}{1-p_{jam}}, \qquad (14)$$

we obtain

$$\ln\frac{p_{jam}}{1-p_{jam}} = -\frac{\rho_{th}}{\Delta}\left.\frac{\partial\langle E\rangle}{\partial p_{jam}}\right|_{\rho}. \quad (15)$$

Substituting the energetic derivative leads to $\ln\frac{p_{jam}}{1-p_{jam}} = -\frac{\gamma}{\eta\Delta}\rho$. Introducing the transition density $\rho_c$ such that $p_{jam}$=1/2 when $\rho = \rho_c$ , the equilibrium condition can be written in centered form

$$\ln\frac{p_{jam}}{1-p_{jam}} = -\frac{\gamma}{\eta\Delta}(\rho-\rho_c). \quad (16)$$

This form ensures that the order parameter varies linearly with the control parameter in the vicinity of the transition, consistent with the empirically observed sigmoidal transition.

Therefore,

$$p_{jam}(\rho) = \frac{p_{max}}{1+\exp\left[-\frac{\gamma(\rho-\rho_c)}{\eta\Delta}\right]}, \quad (17)$$

where $p_{max}$ represents the empirical saturation level of the jam fraction.

Equivalently, defining the symmetric order parameter $\langle m \rangle = 2p_{jam} - 1$, or

$$p_{jam}(\rho) = \frac{p_{max}}{2}\left[1 + \tanh\left(\frac{\gamma(\rho-\rho_c)}{2\eta\Delta}\right)\right] \qquad (18)$$

Up to multiplicative constants absorbed in the definition of the transition width, this expression corresponds to the sigmoidal form used in the main text.

This derivation shows that the sigmoidal transition of the jam fraction emerges naturally from three minimal ingredients:

- a coarse-grained binary description of congestion states,
- an energetic term proportional to normalized network density, and
- a linear dependence of average velocity on the jam fraction.

No assumptions are made about the detailed microscopic distributions of density or velocity; these enter only through renormalized parameters $\rho_c$ and Δ.

The resulting logistic equation therefore represents the equilibrium relation between network density and the macroscopic congestion state.

**Supplementary Note 5: Reconstruction of Free Energy**

Although congestion at the link level is characterized by a continuous variable (volume-over-capacity, VOC), empirical observations reveal two dominant macroscopic regimes corresponding to uncongested and congested network states. Motivated by this bimodal structure, we reconstruct an effective free-energy profile as a function of the network density ρ\rhoρ using the canonical two-state functional form

$$F(\Delta, \rho) = F_0 - \frac{p_{max}}{2}\Delta \ln\left[2cosh\left(\frac{\rho-\rho_c^{(fit)}}{\Delta}\right)\right]. \qquad (19)$$

Here $F_0$ is an additive constant, $\rho_c^{(fit)}$ is the transition density obtained from the sigmoidal fits, and Δ is the empirically determined transition width, corresponding to the effective temperature expressed in density units.

Taking the derivative with respect to ρ yields

$$\frac{\partial F}{\partial \rho} \propto tanh\left(\frac{\rho-\rho c}{\Delta}\right), \qquad (20)$$

which reproduces the sigmoidal dependence of the macroscopic order parameter observed in the data.

The constant $F_0$ does not affect the transition properties and only shifts the potential vertically. Fits were performed using weighted nonlinear least squares together with the binning protocol described above.

*Mobility as Control Parameter (City-Level Analysis)*

At the city scale (Figs. 1–2 in the main text), mobility M is used as the control variable instead of density. Because increased mobility loads the road network, average density is empirically observed to vary monotonically with mobility over the analyzed period. Mobility therefore acts as a natural proxy control parameter for the macroscopic transition.

In this representation, the transition width Δ plays the role of an effective temperature, we called it T, governing the smoothness of the congestion transition. When mobility is expressed in normalized (dimensionless) units, Δ is also dimensionless.

By contrast, in the density-based reconstruction, the transition width retains the units of density (veh/km/lane) and therefore represents a temperature-like parameter defined on the density scale rather than a normalized variable.

After normalizing TCI by its empirical saturation value $TCI_{max}$, $\langle m \rangle = \frac{TCI}{TCI_{max}}$, and centered as $2\langle m \rangle - 1$, making it directly comparable to the canonical two-state expectation.

In this normalized mobility representation, no additional scale factor ($\kappa_{city}$) is required, and the effective temperature is fully determined by the sigmoidal transition width.

*Parameter Roles and Identification*

For clarity, we summarize the estimation status of all parameters introduced in the sigmoidal and free-energy reconstructions:

- $\rho_c$: empirically estimated from sigmoidal fits.
- Δ: estimated from the sigmoidal transition and fixed in the free-energy reconstruction.
- $\kappa_{city}$ : scale factor estimated in free-energy fits (absorbing normalization and sampling effects).
- F0: additive offset without thermodynamic interpretation.

No additional free parameters are introduced beyond those explicitly stated.

Although urban traffic exhibits microscopic interactions (e.g., spillbacks and queue propagation), the present formulation is macroscopic and effective. The free-energy form summarizes the collective outcome of such interactions without modeling explicit pairwise couplings. In this sense, the canonical expression is phenomenological rather than mechanistic.

*Interpretation and limitations*

This noninteracting formulation provides a minimal explanation for the observed sigmoidal mobility–congestion relationship and motivates a thermodynamic interpretation of urban traffic states. Within this framework, mobility reshapes an effective free-energy landscape, and the observed congestion level corresponds to the equilibrium response of the system.

Deviations from this idealized behavior—such as hysteresis, metastability, or history-dependent effects observed in some cities—suggest the presence of interactions between road segments or temporal correlations not captured by the present formulation. The inclusion of explicit interaction terms (e.g., coupling between road segments) would modify the effective potential and could account for staircase-like or hysteretic behaviors observed in specific cases. These extensions remain a natural direction for future development of the framework.

# Supplementary Figures

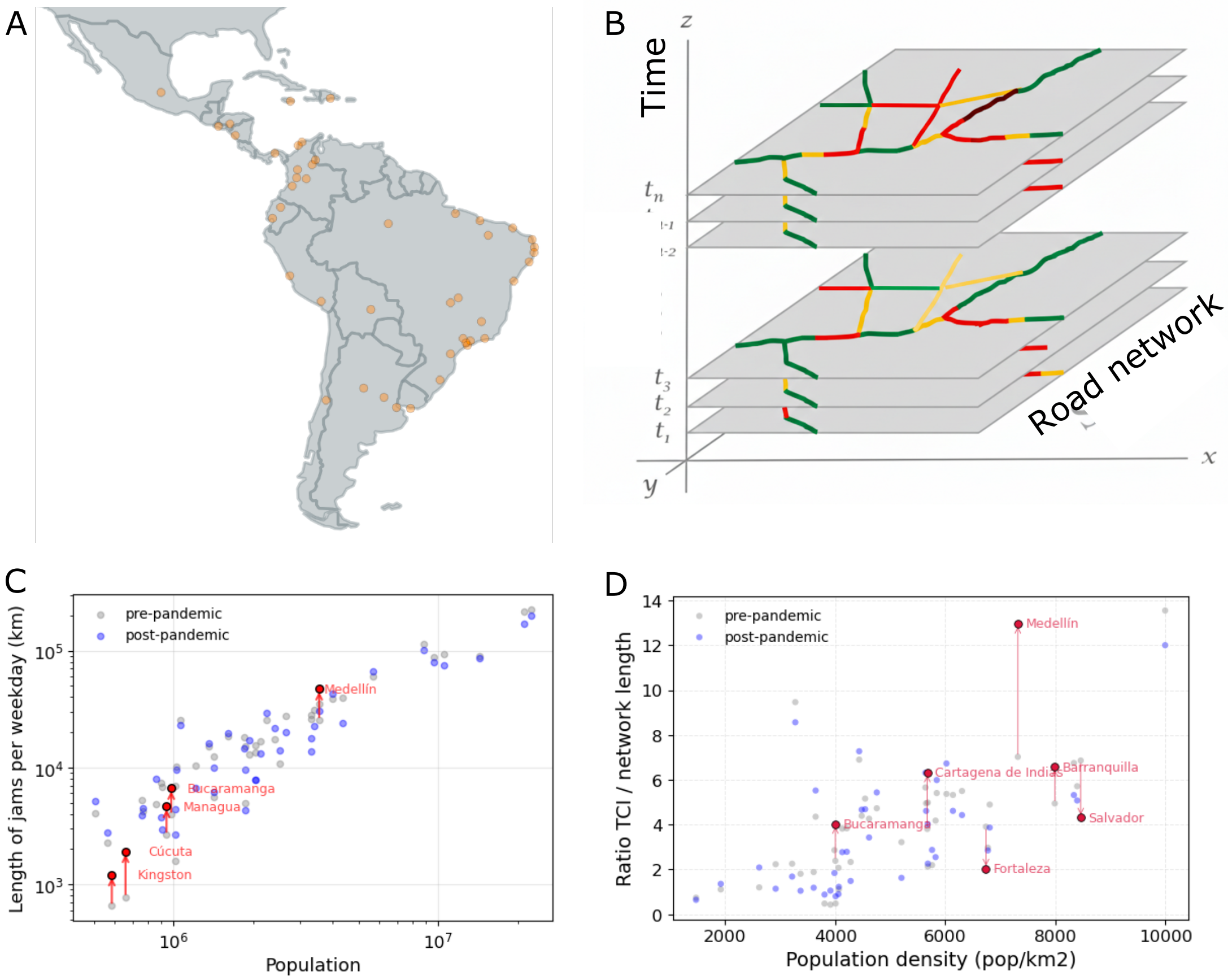

**Supplementary Figure 1. Dataset coverage and construction of the Traffic Congestion Intensity (TCI).** (A) Geographic distribution of metropolitan areas included in the IDB dataset for which Traffic Congestion Intensity (TCI) is computed. Dots indicate cities across Latin America and the Caribbean where sufficient Waze jam-line data are available. (B) Schematic illustration of TCI construction. Every 5 minutes, a network-wide snapshot is taken. Within each snapshot, all identified jam segments are extracted and their lengths summed. The daily TCI corresponds to the time-aggregated jam length across all 5-minute intervals, rescaled relative to the pre-pandemic baseline period. (C) Total daily jam length before and after the pandemic as a function of city population. In both periods, congestion scales with population approximately as a power law, $P^{\alpha}$, consistent with urban scaling behavior. Cities exhibiting the largest post-pandemic deviations are highlighted. (D) Daily jam length normalized by network length as a function of population density. A roughly linear trend emerges, indicating that congestion intensity per unit infrastructure increases proportionally with density.

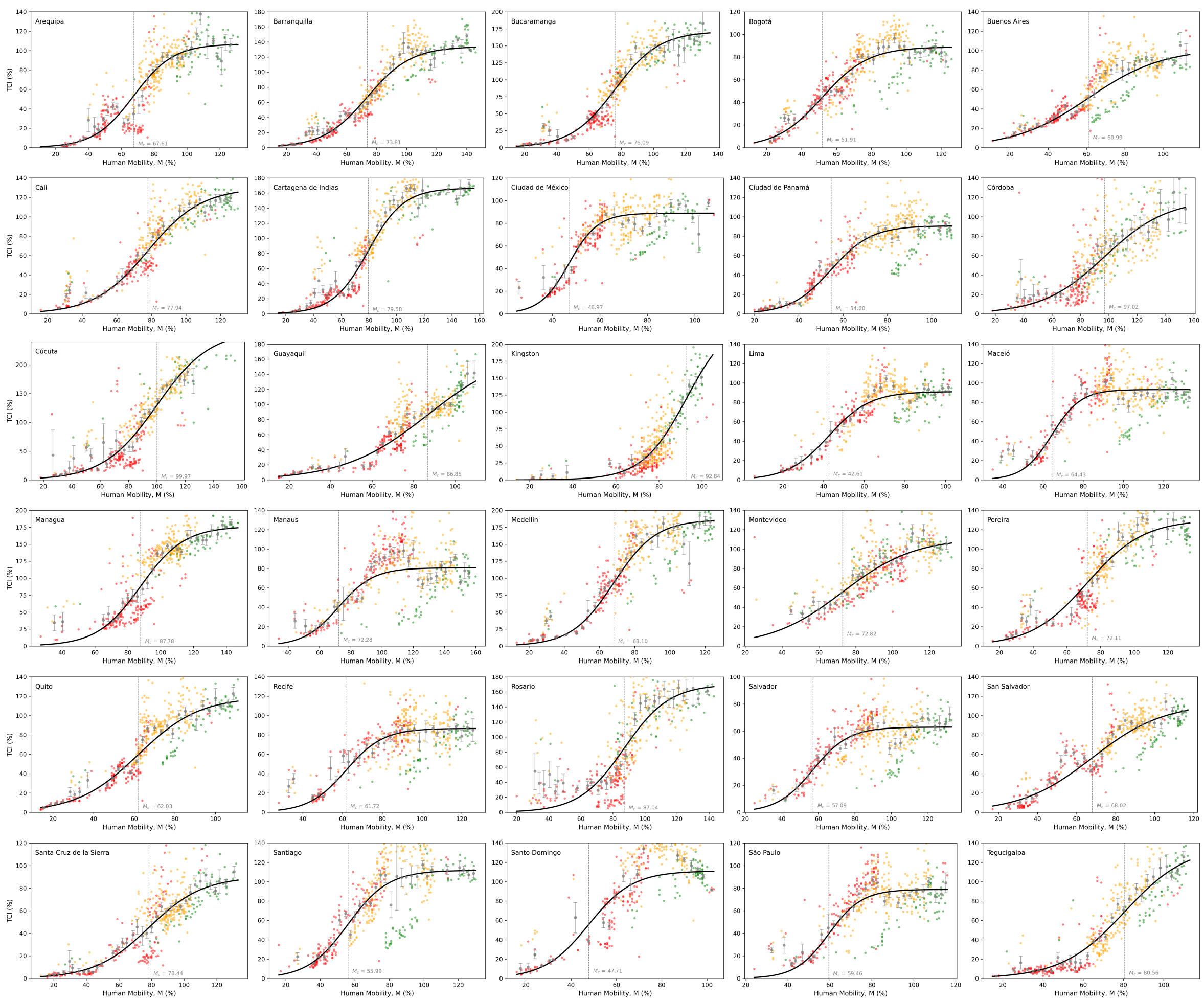

**Supplementary Figure 2. Sigmoidal congestion transitions in daily Traffic Congestion Intensity across cities with high-quality fits.** Each panel shows the daily Traffic Congestion Intensity (TCI) as a function of human mobility for cities with reliable sigmoidal reconstructions **(**31 cities, including Rio de Janeiro in the main text; all with fit quality $R^2 > 0.8$). Points are colored by year (2020 red, 2021 yellow, 2022 green). Solid black curves show the best-fit sigmoidal transition; error bars indicate binned variability. Because the underlying baselines for mobility and congestion differ across metropolitan areas (both in the x- and y-axes), direct cross-city comparison in raw units is not meaningful. This motivates the transformation used in the main text, where mobility is centered as $M - M_c$ and congestion is normalized into the magnetization-like variable $2\langle m \rangle - 1$, enabling collapse and comparison across cities. Several cities (e.g., Kingston, Guayaquil, and Tegucigalpa) do not exhibit a clear saturation plateau within the observed mobility range, suggesting that the system has not yet reached the high-congestion branch during the period analyzed. Notably, despite reported increases in private vehicle use in multiple cities (including Kingston, Medellín, Cartagena, Barranquilla, Cúcuta, Bucaramanga, and Rosario), Kingston still does not display saturation in TCI, consistent with a transition that remains incomplete over the available range. In such cases, T and $M_c$ estimates are less constrained and should be interpreted cautiously.

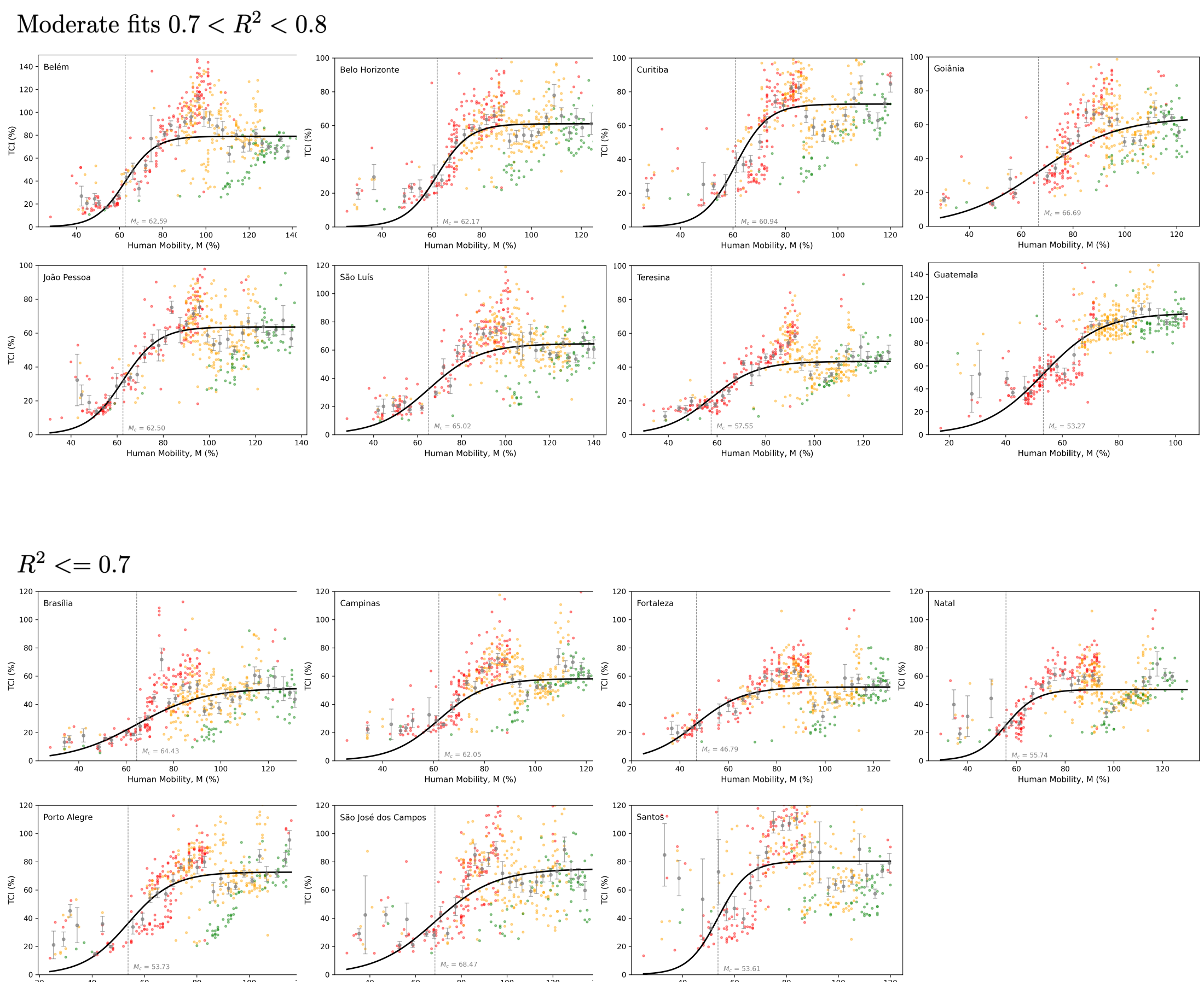

**Supplementary Figure 3. Cities with intermediate and low-quality sigmoidal fits.** Each panel shows daily Traffic Congestion Intensity (TCI) as a function of human mobility for cities where the sigmoidal reconstruction exhibits intermediate or low goodness of fit. Only 7 cities display poor fits, while the majority of cities continue to follow the transition structure described in the main text. Notably, 14 of the 15 lowest-quality fits correspond to Brazilian cities, many of them medium or small urban areas. During the pandemic period, Brazil experienced heterogeneous policy responses and mobility adherence patterns, these deviations coincide temporally with documented heterogeneity in mobility responses during the pandemic. In smaller metropolitan areas, such variability in mobility patterns may weaken the emergence of a well-defined sigmoidal transition, leading to noisier or structurally distorted congestion–mobility relations. Despite these deviations, the overall prevalence of robust sigmoidal fits across the dataset supports the generality of the transition framework.

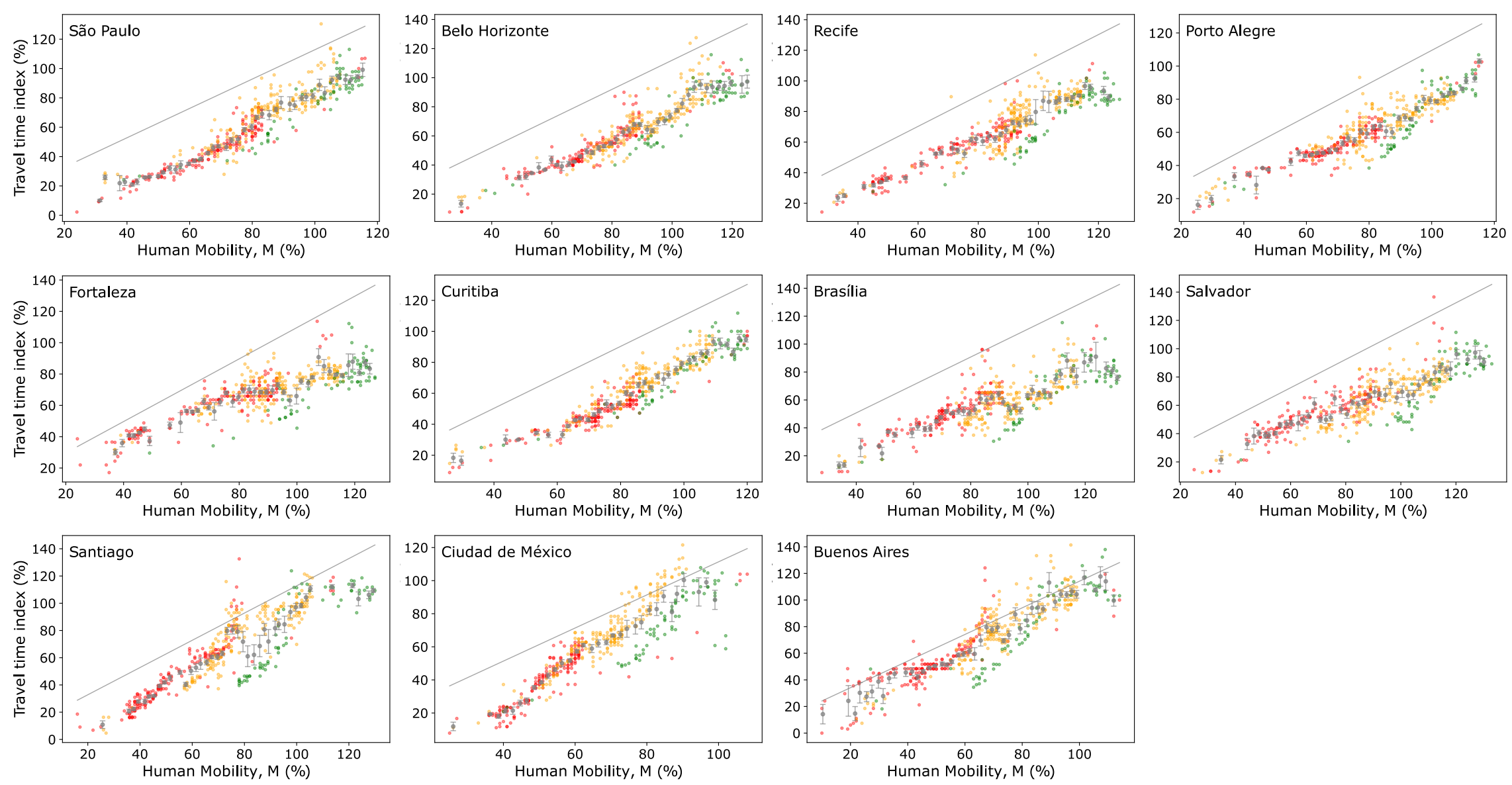

**Supplementary Figure 4. Linear relationship between Travel Time Index and human mobility.** Each panel shows the Travel Time Index (TTI, %) as a function of human mobility M (%) for the cities where TomTom data are available. Points are colored by year (2020 red, 2021 yellow, 2022 green). Across all cities, the relationship is approximately linear with slope close to unity, as indicated by the reference line. This proportionality suggests that, at the aggregate level, increases in mobility translate almost one-to-one into increases in travel time delays.

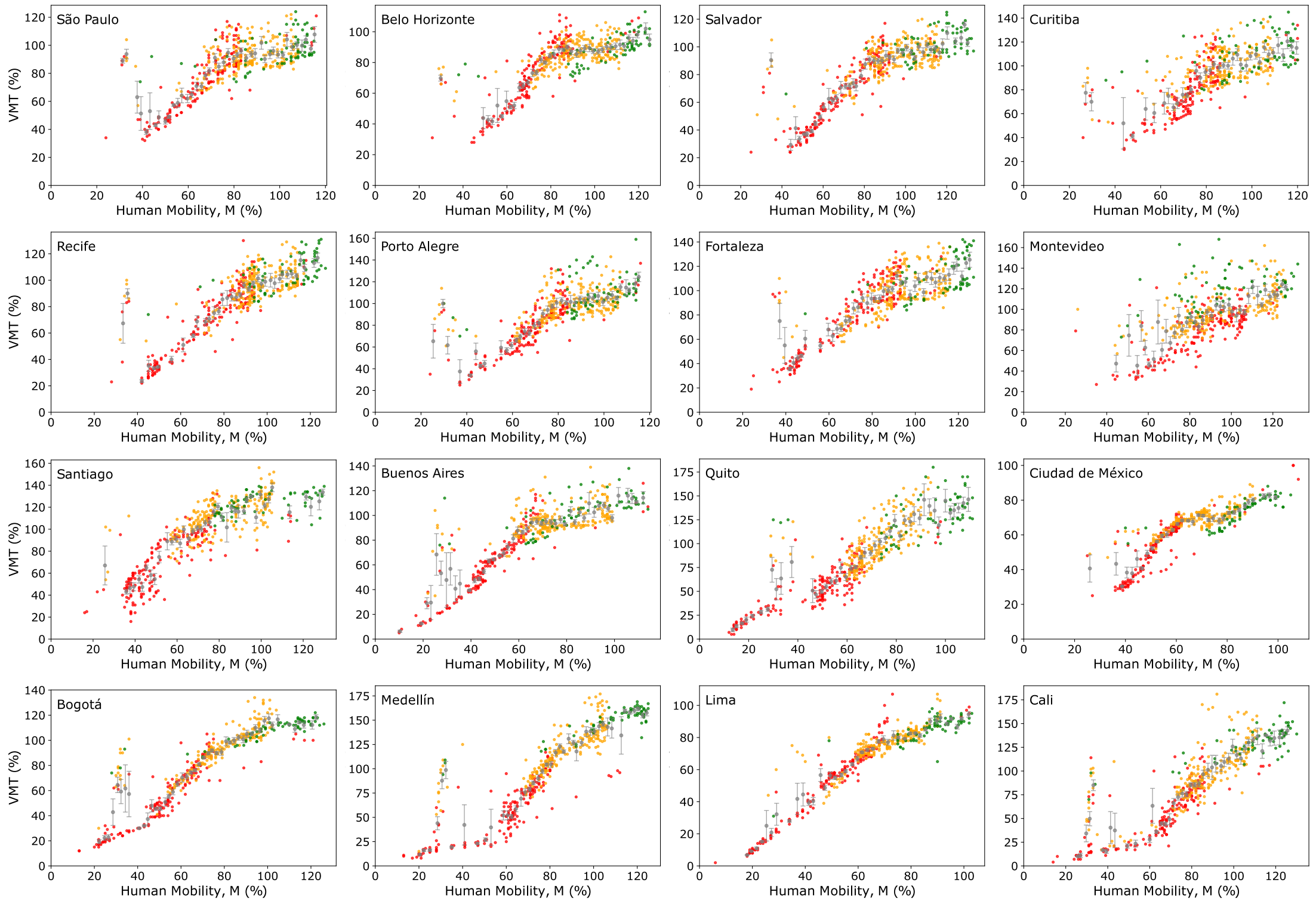

**Supplementary Figure 5. Vehicle Miles Traveled (VMT) as a function of human mobility.** Each panel shows total Vehicle Miles Traveled (VMT, %) versus human mobility M (%) for the cities where VMT data are available. Points are colored by year (2020 red, 2021 yellow, 2022 green). Across metropolitan areas,

VMT exhibits an approximately linear increase with mobility at low-to-intermediate values, followed by a progressive saturation at high mobility levels. This pattern mirrors the behavior reported in the main text, where macroscopic observables grow proportionally with mobility before approaching a capacity-limited regime. The saturation of VMT at high mobility suggests that increases in aggregate movement do not translate indefinitely into proportional increases in traveled distance, consistent with structural constraints imposed by network capacity.

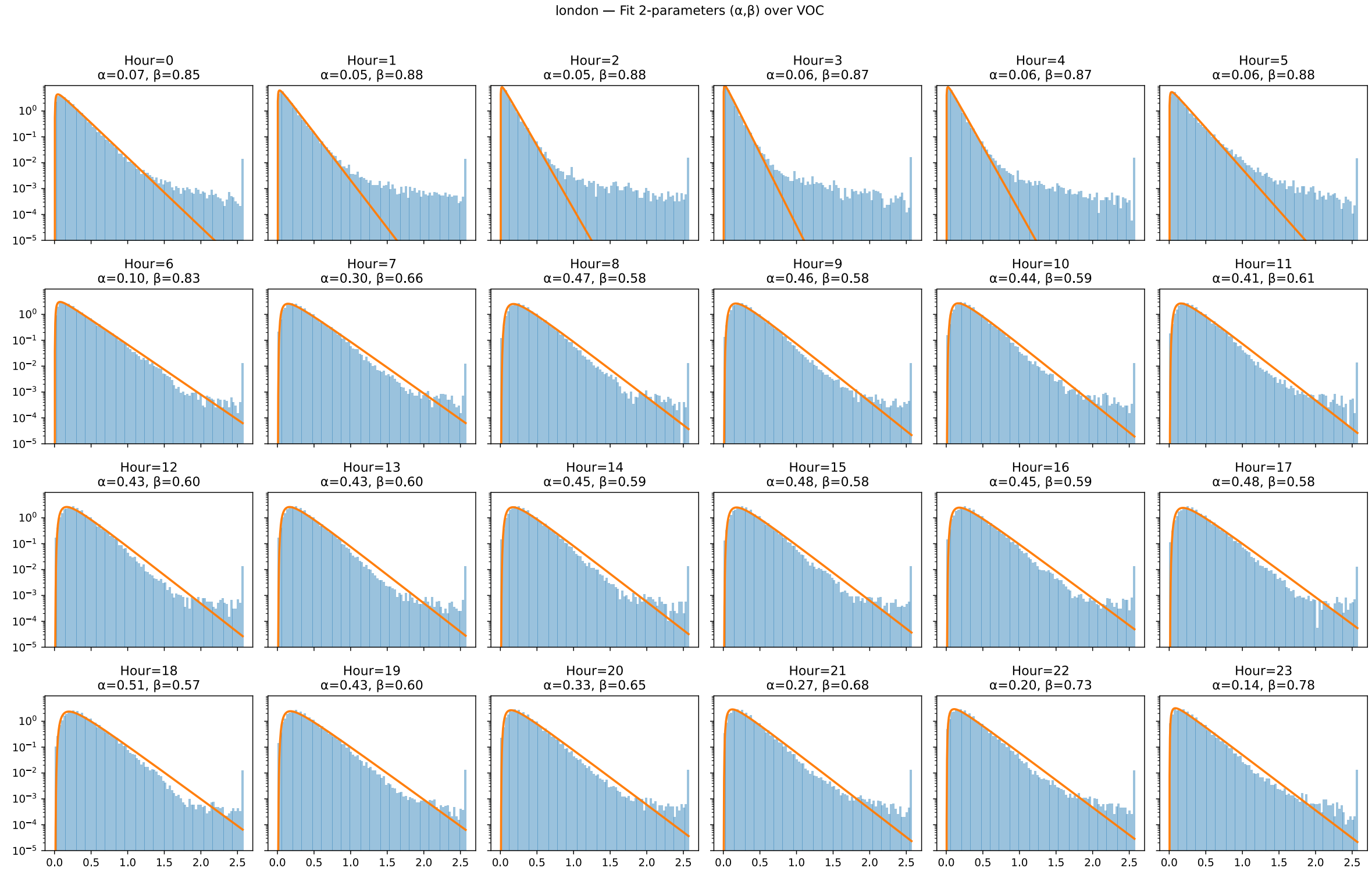

**Supplementary Figure 6 (panel). Hourly fits of the normalized VOC distribution (London example).** Histograms show the empirical distributions of normalized volume-over-capacity $x = \frac{VOC}{\langle VOC \rangle_t}$ for each hour of the day (0–23) in London. Solid curves correspond to two-parameter fits of the form $P_t(\mathrm{x}) \propto exp\,[-\alpha_t/x - \beta_t x]$. The fitted parameters α are indicated in each panel. While α varies systematically over the course of the day, reflecting changes in low-VOC behavior, the parameter β remains comparatively stable, particularly during daytime hours. The approximate constancy of β supports its interpretation as a temperature-like parameter governing the exponential decay of high-VOC states. The consistency of the functional form across all 24 hours indicates that the canonical-like structure of the VOC distribution is not restricted to specific time windows but persists throughout daily traffic cycles. The persistence of the functional form across all hours suggests that the canonical-like structure of the VOC distribution is a robust property of the system, rather than a transient artifact of specific traffic regimes.

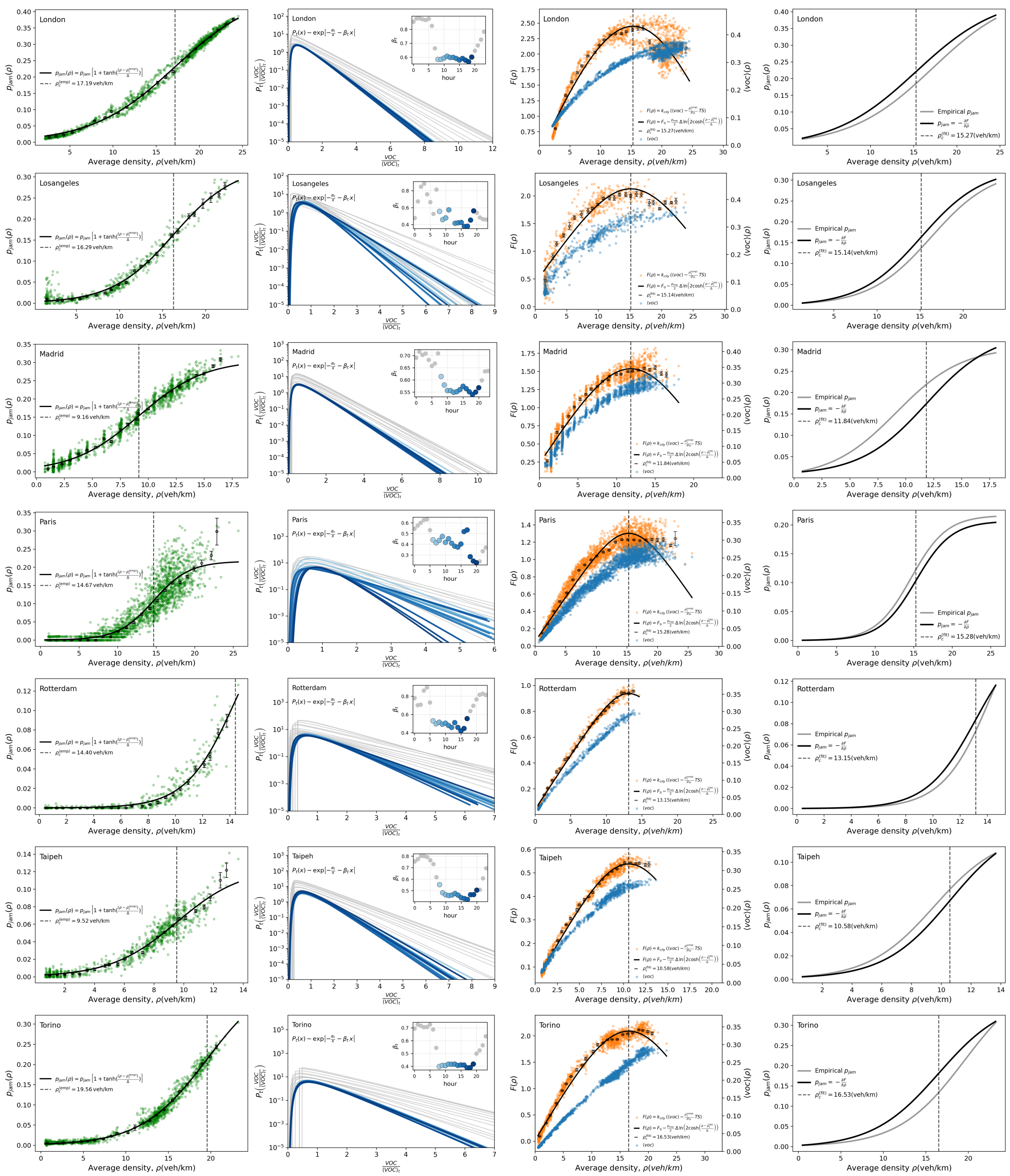

**Supplementary Figure 7. Empirical reconstruction of the thermodynamic congestion framework (full panel).** Each row corresponds to a city. **Left column:** Fraction of jammed links $p_{jam}$ as a function of average network density. Points correspond to aggregated observations; solid curves show the sigmoidal fits used to extract the critical density $\rho_c^{(emp)}$. Vertical dashed lines indicate the inferred transition density. **Second column:** Hourly distributions of normalized volume-over-capacity, $x = \frac{VOC}{\langle VOC \rangle_t}$ . Curves are fitted using the functional form $P_t(\mathrm{x}) \propto exp\left[-\alpha_t/x - \beta_t x\right]$. The inset shows the hourly variation of the decay parameter $\beta_t$, which remains approximately stable throughout the day, supporting the interpretation of $T_{VOC,t} = \frac{1}{\beta_t}$ as an effective temperature-like parameter. **Third column:** Reconstructed Helmholtz free energy $F(\rho) \sim \kappa_{city}\left(\langle VOC \rangle - \frac{\rho_c^{(emp)}}{\rho_{th}} TS\right)$ (orange, left axis) as a function of density. The black curve shows the fit $F(\rho) =$

$F_0 - \frac{p_{max}}{2} \Delta \ln \left[ 2cosh \left( \frac{\rho - \rho_c^{(fit)}}{\Delta} \right) \right]$. Blue points (right axis) display the average $\langle VOC \rangle$ versus density, analogous to a macroscopic fundamental diagram. In most cities, the entropic (Helmholtz) transition occurs at lower density than the peak of $\langle VOC \rangle$, indicating that configurational reorganization emerges before throughput collapse. **Right column:** Sigmoidal congestion curve reconstructed as -∂F/∂ρ, compared with the empirical $p_{jam}$. The agreement confirms that the observed macroscopic transition is consistent with a free-energy description.

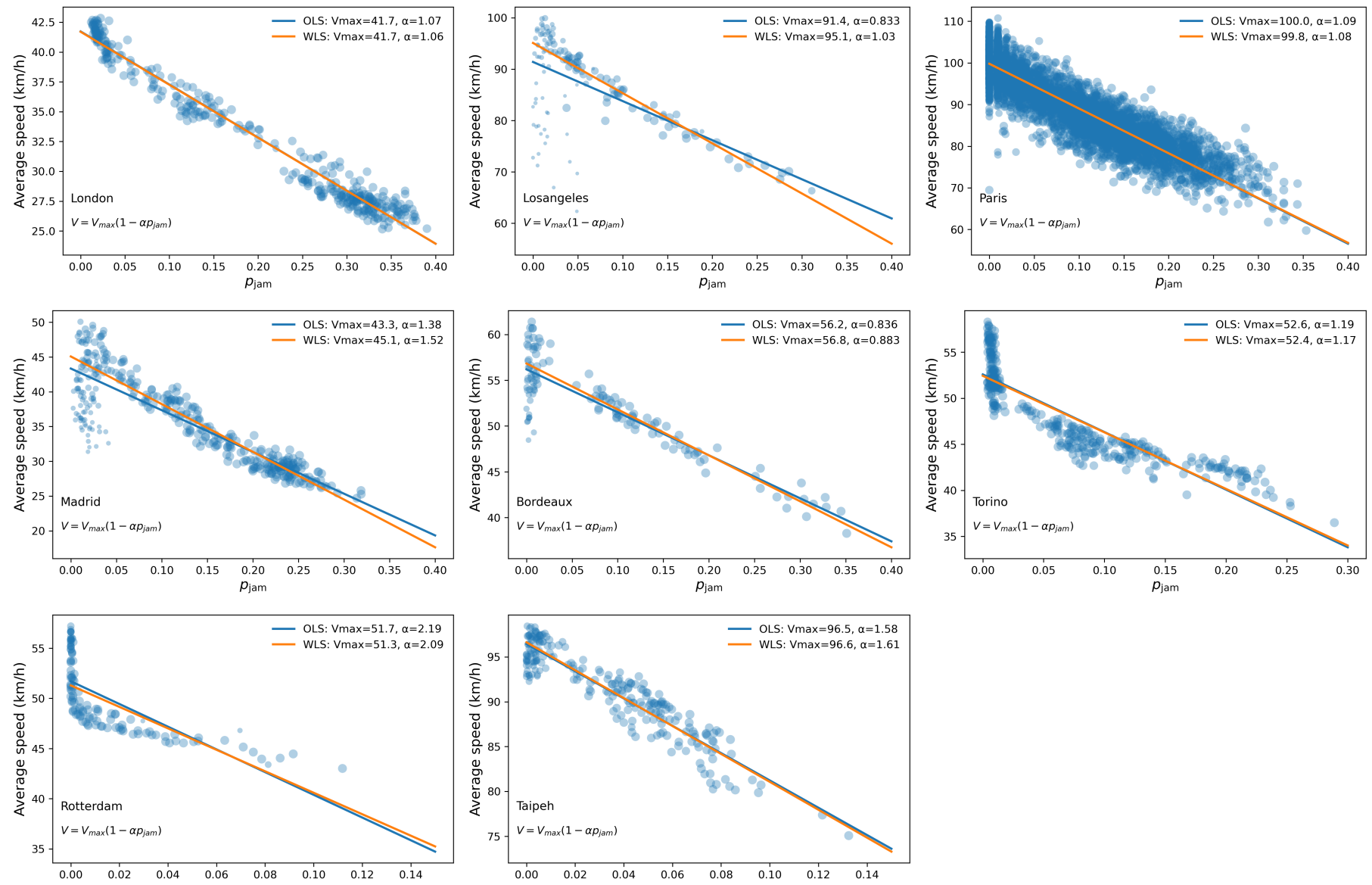

**Supplementary Figure 8: Scatter plots show the relationship between the average network speed v and the fraction of congested links $p_{jam}$ across several cities.** Each point corresponds to an aggregated time interval. A clear approximately linear decrease of speed with increasing congestion is observed. The empirical relationship can be described by $v(\rho, p_{jam}) = v_{max}[1 - \gamma p_{jam}]$ where $v_{max}$ represents the free-flow network speed and $\gamma$ captures the sensitivity of the average speed to the fraction of congested links. This equation resembles the two-fluid formulation of urban traffic introduced by Prigogine and Herman (10)**,** in which the macroscopic speed decreases proportionally to the fraction of vehicles belonging to the congested phase. Ordinary least-squares (OLS) and weighted least-squares (WLS) fits are shown for comparison across cities.

## Supplementary Tables

| City | Country | $p_{\max}^{(emp)}$ | $\rho_c^{(emp)}$ | $\Delta$ | $T_{jam}$ | $\kappa_{city}$ | $p_{\max}^{(fit)}$ | $\rho_c^{(fit)}$ | $F_0$ | $\beta$ | Std($\beta$) | $\alpha$ | Std($\alpha$) |
|---|---|---|---|---|---|---|---|---|---|---|---|---|---|
| Paris | France | 0.11 | 14.67 | 4.63 | 0.32 | 6.0 | 0.10 | 15.28 | 1.63 | 0.40 | 0.10 | 1.36 | 0.76 |
| Los Angeles | US | 0.16 | 16.29 | 7.12 | 0.44 | 8.0 | 0.16 | 15.14 | 2.94 | 0.47 | 0.07 | 0.87 | 0.28 |
| Bordeaux | France | 0.19 | 15.20 | 6.61 | 0.43 | 12.0 | 0.19 | 15.57 | 3.07 | 0.38 | 0.02 | 1.32 | 0.10 |
| Taipei | Taiwan | 0.06 | 9.52 | 4.42 | 0.46 | 2.4 | 0.07 | 10.58 | 0.74 | 0.46 | 0.02 | 0.86 | 0.09 |
| Madrid | Spain | 0.15 | 9.16 | 5.89 | 0.64 | 8.0 | 0.17 | 11.84 | 2.22 | 0.56 | 0.01 | 0.53 | 0.03 |
| Rotterdam | Netherlands | 0.11 | 14.40 | 3.60 | 0.25 | 3.6 | 0.08 | 13.15 | 1.15 | 0.49 | 0.04 | 0.77 | 0.14 |
| London | UK | 0.23 | 17.19 | 9.58 | 0.56 | 13.0 | 0.23 | 15.27 | 3.95 | 0.59 | 0.02 | 0.44 | 0.05 |
| Torino | Italy | 0.21 | 19.56 | 7.79 | 0.40 | 13.0 | 0.18 | 16.53 | 3.08 | 0.41 | 0.01 | 1.12 | 0.06 |

**Supplemementary Table 1.** Summary of fitted parameters across cities.